\documentstyle[12pt,aps,harvard]{revtex}
\begin{document}
\baselineskip=13pt
\renewcommand{\topfraction}{1}
\renewcommand{\bottomfraction}{1}
\renewcommand{\textfraction}{0}
\renewcommand{\floatpagefraction}{1}
\def\Atoday{\ifcase\month\or
  January\or February\or March\or April\or May\or June\or
  July\or August\or September\or October\or November\or December\fi
  \space\number\day, \number\year}
\input psfig
\title{Solar Fusion Cross Sections}
\author{Eric G. Adelberger} 
\address{Nuclear Physics Laboratory, University of Washington, Seattle, WA
98195}
\author{Sam M. Austin}
\address{Department of Physics and Astronomy and NSCL, Michigan 
State University, East Lansing, MI 48824}
\author{John N. Bahcall}
\address{School of Natural Sciences, Institute for Advanced Study, 
Princeton, NJ 08540}
\author{A. B. Balantekin}
\address{Department of Physics, University of Wisconsin, Madison, WI 53706}
\author{Gilles Bogaert}
\address{C.S.N.S.M., IN2P3-CNRS, 91405 Orsay Campus, France}
\author{Lowell S. Brown}
\address{Department of Physics, University of Washington, Seattle, WA 98195}
\author{Lothar Buchmann}
\address{TRIUMF, 4004 Wesbrook Mall, Vancouver, B.C., Canada V6T2A3}
\author{F. Edward Cecil}
\address{Department of Physics, Colorado School of Mines, Golden, CO 80401}
\author{Arthur E. Champagne}
\address{Department of Physics and Astronomy, University of North
Carolina, Chapel Hill NC 27599}
\author{Ludwig de Braeckeleer}
\address{Duke University, Durham, NC 27708}
\author{Charles A. Duba and Steven R. Elliott}
\address{Nuclear Physics Laboratory, University of Washington, Seattle, WA
98195}
\author{Stuart J. Freedman}
\address{Department of Physics, University of California, Berkeley, CA
94720}
\author{Moshe Gai}
\address{Department of Physics U46, University of Connecticut, Storrs,
CT 06269}
\author{G. Goldring}
\address{Department of Particle Physics, Weizmann Institute of Science,
Rehovot 76100, Israel}
\author{Christopher R. Gould}
\address{Physics Department,, North Carolina State University,
Raleigh, NC 27695}
\author{Andrei Gruzinov}
\address{School of Natural Sciences, Institute for Advanced Study, 
Princeton, NJ 08540}
\author{Wick C. Haxton}
\address{Department of Physics, University of Washington, Seattle, WA 98195}
\author{Karsten M. Heeger}
\address{Nuclear Physics Laboratory, University of Washington, Seattle, WA
98195}
\author{Ernest Henley}
\address{Department of Physics, University of Washington, Seattle, WA 98195}
\author{Calvin W. Johnson}
\address{Department of Physics and Astronomy, Louisiana State
University, Baton Rouge, LA 70803}
\author{Marc Kamionkowski}
\address{Physics Department, Columbia University, New York, NY 10027}
\author{Ralph W. Kavanagh and Steven E. Koonin}
\address{California Institute of Technology, Pasadena, CA
91125}
\author{Kuniharu Kubodera}
\address{Department of Physics and Astronomy, University of South
Carolina, Columbia, SC 29208}
\author{Karlheinz Langanke}
\address{University of Aarhus, DK-8000, Aarhus C, Denmark}
\author{Tohru Motobayashi}
\address{Department of Physics, Rikkyo University, Toshima, Tokyo 171, Japan}
\author{Vijay Pandharipande}
\address{Physics Department, University of Illinois, Urbana, IL 61801}
\author{Peter Parker}
\address{Wright Nuclear Structure Laboratory, Yale University, New
Haven, CT 06520}
\author{R. G. H. Robertson}
\address{Nuclear Physics Laboratory, University of Washington, Seattle, WA
98195}
\author{Claus Rolfs}
\address{Experimental Physik III, Ruhr Universit\"at Bochum, D-44780
Bochum, Germany}
\author{R. F. Sawyer}
\address{Physics Department, 
University of California , Santa Barbara, CA 93103}
\author{N. Shaviv}
\address{California Institute of Technology, 130-33, Pasadena, CA 91125}
\author{T. D. Shoppa}
\address{TRIUMF, 4004 Wesbrook Mall, Vancouver, B.C., Canada V6T2A3}
\author{K. A.  Snover and Erik Swanson}
\address{Nuclear Physics Laboratory, University of Washington, Seattle, WA
98195}
\author{Robert E. Tribble}
\address{Cyclotron Institute, Texas A\&M University, College Station,
TX 77843}
\author{Sylvaine Turck-Chi\`eze}
\address{CEA, DSM/DAPNIA, Service d'Astrophysique, CE Saclay, 
91191 Gif-sur-Yvette Cedex, France}
\author{John  F. Wilkerson}
\address{Nuclear Physics Laboratory, University of Washington, Seattle, WA
98195}
\maketitle

\begin{abstract}
We  review and analyze the   available information for 
 nuclear fusion cross sections that are most important for solar
 energy generation and solar neutrino production. We provide best
 values for the low-energy cross-section factors and, wherever
 possible, estimates of the uncertainties. We also describe the most
 important experiments and calculations that are required in order to
 improve our knowledge of solar fusion rates. 
\end{abstract}
\vglue-.25in
\tableofcontents
\vglue-.25in

\section{Introduction}
\label{intro}

This section describes in Sec.~\ref{motive}  
the reasons why a critical analysis of what is
known about solar fusion reactions is timely and important,
summarizes in Sec.~\ref{thiswork} the process by  which this collective
manuscript  was
written , and provides 
 in Sec.~\ref{contents} a brief outline of the structure of the paper.

\subsection{Motivation}
\label{motive}

The original motivation of solar neutrino experiments was to use the
neutrinos ``..to see into the interior of a star and thus verify
directly the hypothesis of nuclear energy generation in stars''
(Bahcall, 1964; Davis, 1964).\nocite{Bahcall64,Davis64} This goal 
has now been achieved by four
pioneering experiments: Homestake (Davis,  1994),\nocite{Davis94}
Kamiokande (Fukuda {\it et al.}, 1996),\nocite{Fukuda96} 
 GALLEX (Kirsten {\it et al.},  1997),\nocite{Kirsten96} and 
SAGE (Gavrin {\it et al.},  1997).\nocite{Gavrin96} These 
experiments provide direct evidence 
that the stars shine and evolve as the result of nuclear fusion
reactions among light elements in their interiors.

Stimulated in large part by the precision obtainable in solar neutrino
experiments and by solar neutrino calculations with standard models of
the sun, 
our knowledge of the 
low-energy cross sections for fusion reactions among light
elements has been greatly refined by many hundreds of careful studies of the
rates of these reactions. The rate of progress was particularly
dramatic in the first few years following the proposal of the chlorine
(Homestake) 
experiment in 1964.

In 1964, when the chlorine solar neutrino experiment was
proposed (Davis, 1964; Bahcall, 1964),\nocite{Davis64,Bahcall64} 
the rate of the ${\rm^3He}$-${\rm^3He}$
reaction was estimated (Good, Kunz, and Moak, 1954; Parker, Bahcall, and Fowler,
1964)\nocite{Good54,parker64}  
 to be $5$ times slower 
than the current best estimate and the uncertainty in the low-energy
cross section was
estimated (Parker, Bahcall, and Fowler,  1964)\nocite{parker64}  
to be ``as much as a factor of $5$ or $10$.''  Since the 
${\rm^3He}$-${\rm^3He}$ reaction competes with the 
${\rm^3He}$-${\rm^4He}$ 
reaction---which leads to high energy neutrinos---the
calculated fluxes for the higher energy neutrinos were overestimated in
the earliest days of solar neutrino research.
The most significant uncertainties, in the rates of the 
${\rm^3He}$-${\rm^3He}$,
the ${\rm^3He}$-${\rm^4He}$, and the ${\rm^7Be}$-$p$ reactions, were 
much reduced after
just a few years of intensive experimental research in the middle and late
1960s (Bahcall and Davis, 1982).\nocite{Bahcalldavis}

Over the past three decades, steady and impressive progress has been
made in refining the rates of these and other reactions that produce
solar energy and solar neutrinos. (For reviews of previous work on
this subject, see, e.g., Fowler, Caughlan, and Zimmerman, 1967, 1975;
Bahcall and Davis, 1982; Clayton, 1983; Fowler, 1984; Parker, 1986;
Rolfs and Rodney, 1988; Caughlan and Fowler, 1988; Bahcall and
Pinsonneault, 1992, 1995; Parker and Rolfs, 1991).
\nocite{Fowler67,Fowler75,Bahcalldavis,Clayton,Fowler84,Parker86,Rolfs,caughlan88,pinson,pinson95,Parker91}
An independent assessment of nuclear fusion reaction rates is being
conducted by the European Nuclear Astrophysics Compilation of Reaction
Rates (NACRE) (see, e.g., Angulo, 1997);\nocite{angulo97} the 
results from this compilation,
which has  broader goals  than our study and in particular 
does not focus on precision solar rates,  are
not yet available.

However, an unexpected development has occurred. The accuracy of the
solar neutrino experiments and the precision of the theoretical
predictions based upon standard solar models and standard electroweak
theory have made possible extraordinarily sensitive tests of new
physics, of physics
beyond the minimal standard electroweak model.
Even more surprising is the fact that, for the past three
decades,  the neutrino experiments
have consistently disagreed with standard predictions, despite
concerted efforts by many 
physicists, chemists, astronomers, and engineers
to find ways out of this dilemma.

The four pioneering solar neutrino experiments together provide
evidence for  physics beyond the standard electroweak theory. The
Kamiokande (Fukuda {\it et al.}, 1996)\nocite{Fukuda96} 
and the chlorine (Davis, 1994)\nocite{Davis94} experiments
appear to be inconsistent with
each other if nothing happens to the neutrinos after they are created
in the center of the sun (Bahcall and Bethe, 1990).\nocite{bethe90}
Moreover, the well calibrated gallium solar
neutrino experiments GALLEX (Kirsten {\it et al.},  1997),\nocite{Kirsten96} and 
SAGE (Gavrin {\it et al.}, 1997)\nocite{Gavrin96} 
are interpreted, if neutrinos do not oscillate or
otherwise change their states on the way to the earth from the solar
core, as indicating an almost complete absence of  $^7$Be neutrinos.
However, we know [see discussion
of Eq.~(\ref{Beratio}) in Sec.~\ref{beelectron}]
that the $^7$Be neutrinos must be present, if there is no new
electroweak physics occurring, because of the demonstration that
$^8$B neutrinos are observed by the Kamiokande
solar neutrino experiment. Both $^7$Be and $^8$B neutrinos are
produced by capture on $^7$Be ions.

New solar neutrino 
experiments are currently underway to test for evidence of new
physics with exquisitely precise and sensitive techniques.  These
experiments include a huge pure water Cerenkov detector known as
Super-Kamiokande (Suzuki, 1994; Totsuka, 1996),\nocite{Suzuki94,Totsuka96}  
a kiloton of heavy water, SNO, that will
study both neutral and charged currents (Ewan {\it et al.}, 1987,
1989; McDonald, 1995),\nocite{Ewan87,Ewan89,Mcdonald95}
a large organic scintillator, BOREXINO, that will investigate
lower energy neutrinos than has previously been possible (Arpesella
{\it et al.}, 1992; Raghavan, 1995),\nocite{Arpesella92,Raghavan95} 
and a $600$ ton liquid argon time projection chamber, ICARUS,
that will provide detailed information on the surviving $^8$B $\nu_e$
flux (Rubbia, 1996; ICARUS collaboration, 1995; Bahcall {\it et al.}, 1986).
\nocite{Rubbia,Icarus95,Bahcall86} 
With these new detectors, it will be possible to search for evidence
of new physics that is independent of details of solar model
predictions. [Discussions of solar neutrino experiments and the related
physics and astronomy can be found at, for example,
http://www.hep.anl.gov/NDK/Hypertext/nuindustry.html,
http://neutrino.pc.helsinki.fi/neutrino/,  and
http://www.sns.ias.edu/~jnb .]

However, our ability to interpret the existing and new solar neutrino
experiments is limited by the imprecision in our knowledge of the 
relevant nuclear fusion cross sections.  To cite the most important
example, 
the calculated rate of events in the Super-Kamiokande and SNO solar
neutrino experiments is directly proportional to the 
 rate measured in the laboratory at low energies 
for the ${\rm ^7Be}(p,\gamma){\rm ^8B}$ reaction.  
This reaction is so rare in the
sun, that the assumed rate of  ${\rm ^7Be}(p,\gamma){\rm ^8B}$ 
has only a negligible
effect on solar models and therefore on the structure of the sun.  
The predicted rate of neutrino events in the  interval $2$ MeV to $15$ MeV
is directly proportional to the measured laboratory
rate of the ${\rm ^7Be}(p,\gamma){\rm ^8B}$ reaction. Unfortunately, the 
low-energy cross-section factor for the production of $^8$B is the least
well known of the important cross sections in the $pp$ chain.

We will concentrate in this review on the low-energy cross section
factors, $S$, that determine the rates for the most important solar
fusion reactions. The local rate of a non-resonant fusion reaction
can be written in the following form (see, e.g., Bahcall,
1989):\nocite{bahcall89} 

\begin{equation}
 \left<\sigma v\right> = 1.3005 \times 10^{-15} \left[Z_1 Z_2 \over A T_6^2
\right ]^{1/3} f S_{\rm eff} ~ \exp\left(-\tau \right) \rm~cm^{3} ~ s^{-1}.
\label{rateeqn}
\end{equation}
Here $Z_1$, $Z_2$ are the nuclear charges of the fusing ions, 
$A_1$,$A_2$ are the atomic mass numbers, 
$A$ the reduced mass $A_1 A_2/(A_1 + A_2)$, 
 $T_6$ is the temperature
in units of $10^{6}$~K, and the cross-section factor $S_{\rm eff}$
(defined below) is in keV~b. The most probable energy, $E_0$, at which the
reaction occurs is 

\begin{eqnarray}
E_0 & = &\left[(\pi \alpha Z_1 Z_2 kT)^2 (m A c^2/2) \right]^{1/3}\nonumber\\
& = &1.2204(Z_1^2  Z_2^2 A T_6^2)^{1/3} ~{\rm keV}.
\label{defnezero}
\end{eqnarray}
The energy $E_0$ is also known as the Gamow energy.
The exponent  $\tau$ that occurs in Eq.~(\ref{rateeqn}) dominates the
temperature dependence of the reaction rate and is given by 

\begin{equation}
\tau = 3 E_0/{kT} = 42.487
\left(Z_1^2 Z_2^2 A T_6^{-1} \right)^{1/3} .
\label{defntau}
\end{equation}
For all the important reactions of interest in solar fusion, $\tau$ is
in the range $15$ to $40$. The quantity $f$ is a correction factor due
to screening first calculated by Salpeter (1954)\nocite{salpeter54} 
and discussed in this
paper in Sec.~\ref{stellarscreening}.  
The quantity $S_{eff}$ is the effective cross section factor for the
fusion reaction of interest and is evaluated at the most probable
interaction energy, $E_0$.  To first order in $\tau^{-1} $ (Bahcall,
1966),\nocite{bahcallftau}
\begin{equation}
S_{\rm eff} = S(E_0) \left\{ 1 + \tau ^{-1} \left[{5 \over 12} + {5 S^\prime
E_0 \over 2 S } + {S^{\prime \prime }E_0^2 \over S} \right]_{E = E_0}
\right\}.
\label{seffdefn}
\end{equation}
Here $S^\prime = {\rm d}S/{\rm d}E $.
In most analyses in the literature, the values of $S$ and
associated derivatives are quoted at zero energy, not at $E_0$.
In order to relate (\ref{seffdefn}) to the usual formulae, one must
express the relevant quantities in terms of their values at $E ~=~ 0$.
The appropriate connection is

\begin{eqnarray}
S_{\rm eff} (E_0) &\simeq& \hfill\nonumber\\
S(0) \left[1 \right.&+&\left. {5 \over 12 \tau} + {S^\prime \left(E_0 + {35
\over 36} kT \right) \over S }
+ {S^{\prime \prime} E_0 \over S} \left({E_0 \over 2}
+ {89 \over 72}kT \right) \right]_{E ~=~ 0}.
\label{ssubzero}
\end{eqnarray}
In some contexts, $S_{\rm eff}(E_0)$ is referred to as simply the
`$S$-factor' or `the low-energy $S$-factor'.

For standard solar models (cf. Bahcall, 1989)\nocite{bahcall89}, 
the fusion energy and the $pp$ neutrino flux are generated over a
rather wide range of temperatures,
$8 < T_6 < 16$. The other important fusion reactions and neutrino
fluxes are generated over a more narrow range of physical
conditions. The $^8$B neutrino flux is created in the most restricted
temperature range, $13 < T_6 < 16$ .  The mass density 
(in ${\rm g~cm^{-3}}$) is given
approximately by the relation $\rho = 0.04 T_6^3$ in the temperature
range of interest.

The approximate dependences of the  
solar neutrino fluxes on the
different low-energy nuclear cross-section factors can be calculated for
standard solar models.  
The most important fluxes for solar neutrino experiments that have been
carried out so far, or which are currently being constructed, are the 
low energy neutrinos from the fundamental  $pp$ reaction, $\phi({\rm pp})$,
the intermediate energy $^7$Be line neutrinos, $\phi({\rm ^7Be})$, and
the rare high-energy neutrinos from $^8$B decay, $\phi({\rm ^8B})$.  
The $pp$ neutrinos are the most abundant experimentally-accessible
solar neutrinos and the $^8$B neutrinos have the smallest detectable
flux, according to the predictions of standard models (Bahcall,
1989)\nocite{bahcall89}.

Let $S_{11}$, $S_{33}$, and
$S_{34}$ be the low-energy, nuclear
 cross-section factors 
(defined in Sec.~\ref{extrapolation})
for the $pp$, ${\rm
^3He} + {\rm ^3He}$, and ${\rm ^3He} + {\rm ^4He}$ reactions and let
$ S_{17}$ and $S_{e^-\, 7}$ be the cross-section factors
for the capture by $^7$Be of, respectively, protons and electrons.
Then (Bahcall 1989)\nocite{bahcall89}

\begin{mathletters}
\begin{equation}
\phi (pp) \propto S^{0.14}_{11}\, S^{0.03}_{33}\, S^{-0.06}_{34}\  ,
\label{phipp}
\end{equation}
\begin{equation}
\phi({\rm ^7Be}) \propto S^{-0.97}_{11}\, S^{-0.43}_{33}\,
S^{0.86}_{34}\ ,
\label{phi7be}
\end{equation}
and
\begin{equation}
\phi({\rm ^8B}) \propto S^{-2.6}_{11}\, S^{-0.40}_{33}\,
S^{0.81}_{34}\, S^{1.0}_{17}\, S^{-1.0}_{e^-\, 7}\ .
\label{phi8b}
\end{equation}
\label{lowecross}
\end{mathletters}

Nuclear fusion reactions among light elements both generate solar
energy and produce solar neutrinos.  Therefore, the observed solar
luminosity places a strong constraint on the current  rate of solar
neutrino generation calculated with standard solar models.  
In addition, the shape of the neutrino energy
spectrum from each neutrino source is unaffected, to experimental
accuracy, by the solar environment.  
A good fit to the results from current solar neutrino experiments is
not possible, independent of other, more model-dependent solar issues
provided  nothing happens to the neutrinos
after they are created in the sun (see, e.g., 
Castellani {\it et al.} 1997; Heeger and Robertson,
1996; Bahcall, 1996; Hata, Bludman, and Langacker, 1994, and 
references therein).
\nocite{Cast97,Heeger96,Hata94,Bahcall96}

But, the ultimate limit of our ability to extract astronomical
information and to infer neutrino parameters will be constrained  by our
knowledge of the spectrum of neutrinos created in the center of the 
sun.  Returning to the example of the $^8$B neutrinos, the total flux
(independent of flavor) 
of these neutrinos will be measured in the neutral current experiment
of SNO, and--using the charged current measurements of  SNO and ICARUS--in
Super-Kamiokande.   
This total flux is very sensitive to
temperature, $\phi({\rm ^8B}) \sim S_{17}T^{24}$ (Bahcall and Ulmer, 1996),
\nocite{BahcallU} where $T$ is the central temperature of the sun.
Therefore, our ability to test solar model calculations of the central
temperature profile of the sun is limited by our knowledge of
$S_{17}$.

Existing or planned solar neutrino experiments 
are expected to determine whether the energy spectrum of electron type
neutrinos created in the center of the sun is modified by physics
beyond standard electroweak theory. Moreover, these experiments have
the capability of determining the mechanism, if any, by which new
physics is manifested in solar neutrino experiments  
and thereby determining how the original neutrino
spectrum is altered by the new physics. 
Once we reach this stage, solar neutrino experiments will provide
precision tests of solar model predictions for the rates at which
nuclear reactions occur in the sun.

After the neutrino physics is understood, neutrino experiments will
determine 
the average ratio in the solar interior of the
$^3$He-$^3$He reaction rate to the rate of the 
$^3$He-$^4$He reaction.   This solar ratio of reaction rates,
$R_{33}/R_{34}$,  
can be  inferred directly from the measured total
flux of $^7$Be and $pp$ neutrinos (Bahcall, 1989).\nocite{bahcall89}  
The comparison of the measured
and the calculated ratio of 
$R_{33}/R_{34}$ will constitute a stringent and informative test of
the theory of stellar interiors and nuclear energy generation.
In order to extract the inherent information about the solar interior
from the measured ratio, we must know the nuclear fusion 
cross sections that determine the branching ratios among the different
reactions in the $pp$ chain.

\subsection{The origin of this work}
\label{thiswork}

This paper originated from our joint efforts to critically assess
the state of the nuclear physics important to the solar
neutrino problem.  There are two motivations for taking on 
such a task at this time.  First, we have entered a period
where the sun, and  solar models, can be probed with
unprecedented precision through neutrino flux measurements
and helioseismology.  It is therefore important to assess how 
uncertainties in our understanding of the underlying nuclear
physics might affect our interpretation of such precise measurements.
Second, as the importance of
the solar neutrino problem to particle physics and astrophysics
has grown, so has also the size of the community interested in this
problem.  Many of the interested 
physicists are unfamiliar with the decades of
effort that have been invested in extracting the needed
nuclear reaction cross sections, and thus uncertain about the
quality of the results.  The second goal of this paper
is to provide a critical assessment of the current state of
solar fusion research, describing  what is known
while also delineating the  possibilities  for further
reducing nuclear cross-section uncertainties.

In order to achieve these goals, an international 
 collection of experts on nuclear
physics and solar fusion---representing every speciality (experimental
and theoretical) and every point of view (often conflicting points of
view)---met in a workshop on ``Solar Fusion Reactions.''  
In particular,
the  participants included experts on all the major controversial issues 
discussed in  widely circulated
 preprints or in the published literature. 
The
 workshop was held at the Institute for Nuclear Theory, University of 
Washington, February 17-20, 1997.\footnote{
The workshop was proposed by John Bahcall,
the principal editor of this paper, in a
letter submitted to the Advisory Committee of the Institute for Nuclear
Theory, August 20, 1996.  W. Haxton, P. Parker, and H. Robertson served
as joint organizers (with Bahcall) 
of the workshop and as co-editors of this paper.
All of the co-authors participated actively in some stage of the work
and/or 
the writing of this paper.  We attempted to be complete in our review
 of the literature prior to the workshop meeting and have taken
 account of the most relevant work that has been published prior to
 the submission of this paper in September, 1997.
} 
The goal of the workshop
 was to initiate critical discussions evaluating 
all of the existing measurements and calculations
 relating to solar fusion and to recommend a set of standard 
 parameters and their associated uncertainties on which all of the
 participants could agree. To achieve this goal, 
we undertook  {\it ab initio} analyses of each of the important solar
 fusion reactions; previously cited reviews largely concentrated on
 incremental improvements on earlier work.
This paper is our joint work and represents
 the planned culmination of
 the workshop activities.


At the workshop, we held plenary sessions on each of the
important reactions and also intensive specialized discussions in
smaller groups.  The discussions were led by the following
individuals: extrapolations (K. Langanke), electron screening
(S. Koonin),
$pp$ (M. Kamionkowski), ${\rm ^3He + ^3He}$ (C. Rolfs),
${\rm ^3He + ^4He}$ (P. Parker), $e^- ~+~^7$Be (J. Bahcall),
$p ~+~^7$Be (E. Adelberger), and CNO (H. Robertson).
Initial drafts of each of the sections in this paper were written by
the discussion leaders and their close collaborators.  
Successive iterations of the paper were posted on the Internet so that
they could be read and commented on by each member of the
collaboration,
resulting in an almost infinite number of iterations.
Each section of the paper was reviewed
extensively and critically by co-authors who did not draft that section,
and, in a few cases, vetted by outside experts.

\subsection{Contents}
\label{contents}

The organization of this paper reflects the organization of our
workshop. Section~II describes the theoretical justification and the 
phenomenological situation regarding 
extrapolations from higher laboratory energies to lower solar
energies,
as well as the effects of electron screening 
on laboratory and solar fusion rates.
Sections~\ref{pp-pep}--\ref{cno} contain detailed descriptions of the current
situation with regard to the most important solar fusion reactions.
We do not consider explicitly in this review  
the reactions ${\rm ^2H}(p,\gamma)^3$He, 
${\rm ^7Li}(p,\alpha){\rm ^4He}$, and 
${\rm ^8B}(\beta^+ \nu_e){\rm ^8Be}$, which occur in the
$pp$ chain but whose rates are so fast that the precise cross section
or decay time does not affect the energy generation or the neutrino
flux calculations.  We concentrate our discussion on those reactions
that are most important for calculating solar 
neutrino fluxes or energy production.

In our discussions at the workshop, and in the many iterations that
have followed over the subsequent months, we placed as much emphasis on
determining reliable error estimates as on specifying the best values.
We recognize that, for applications to astronomy and to neutrino
physics,  it is as important to know the limits of our
knowledge as it is to record the preferred cross-section
factors. Wherever possible, 
experimental results are given with $1\sigma$ error bars (unless
specifically noted otherwise). For a few quantities, we have also
quoted estimates of a less precisely defined quantity that we refer to
as an ``effective $3\sigma$'' error (or a maximum likely uncertainty).
In order to meet the challenges and opportunities provided by
increasingly precise solar neutrino and helioseismological data, we
have emphasized in each of the sections on individual reactions the
most important measurements and calculations to be made in the future.

The sections on individual reactions, \ref{pp-pep}--\ref{cno}, 
answer the questions:
``What?'',  ``How Well?'', and ``What Next?''.
Table~\ref{summarytable} summarizes the answers to the questions ``What?''
and ``How Well?''; 
this table gives the best estimates and uncertainties for
each of the principal solar fusion reactions that are discussed in
greater detail later in this paper. 
The different answers to the question ``What Next?'' are given in the
individual Secs.~\ref{extrapolationscreening}--\ref{cno}.

\section{Extrapolation and Screening}
\label{extrapolationscreening}

\subsection{Phenomenological Extrapolation}
\label{extrapolation}


Nuclear fusion reactions occur via a short-range (less than or
comparable to a few fm) strong interaction.  
However, at the
low energies typical of solar fusion reactions 
($\sim 5$ keV to $30$ keV),
the two nuclei must
overcome a sizeable barrier provided by the long-range Coulomb
repulsion before they can come close enough to fuse.  
Therefore,
the energy dependence of a (nonresonant) fusion cross
section is conveniently written in terms of an $S$-factor 
which is defined by the following relation:
\begin{equation}
     \sigma(E) = {S(E) \over E} {\rm exp} \left\{ -2 \pi \eta(E) \right\},
\protect\label{eqsfactor}
\end{equation}
where 
\begin{equation}
     \eta(E) = \frac{Z_1 Z_2 e^2}{\hbar v}
\protect\label{eqsommerfeld}
\end{equation}
is the Sommerfeld parameter.  Here, $E$ is the center-of-mass energy; 
$v=(2E/\mu)^{1/2}$ is the
relative velocity in the entrance channel; $Z_1$ and $Z_2$ are
the charge numbers of the colliding nuclei; $\mu=m A_1
A_2/(A_1+A_2)$ is the reduced mass of the system; $m$ is the
 atomic mass unit; and $A_1$ and $A_2$ are the masses (in units of $m$) of
the reacting nuclei.

The exponential in Eq.~(\ref{eqsfactor}) (the Gamow penetration factor)
takes into account quantum-mechanical tunneling through the
Coulomb barrier; the exponential describes well the rapid decrease of
the cross section with decreasing energy. The Gamow penetration factor
dominates the energy dependence, derived in the WKB approximation, of the
cross section in the low-energy limit.  In the
low-energy regime in which the WKB approximation is valid, the
function $S(E)$ is slowly varying (except for resonances)
and may be approximated by
\begin{equation}
     S(E) \simeq S(0) + S^\prime(0)E + {1\over2} S^{\prime\prime}(0)E^2.
\protect\label{quadapprox}
\end{equation}
The coefficients in Eq.~(\ref{quadapprox})
can often be  determined by  fitting a quadratic formula to laboratory
measurements or theoretical calculations 
of the cross section made at energies of order 100 keV
to several MeV.  The cross section is then extrapolated to
energies, ${\cal O}(10\, {\rm keV})$, typical of solar
reactions, through Eq. (\ref{eqsfactor}).  However, special care has
to be exercised for certain reactions like $^7$Be(p,$\gamma$)$^8$B,
where the $S$-factor at very low energies expected from theoretical
considerations 
cannot be seen in  available data (cf. discussion in
Sec.~\ref{bep}). 

The WKB approximation for the Gamow penetration factor is 
valid if the argument of the exponential is large, $2
\pi \eta \gtrsim 1$.  This condition is 
satisfied for the energies over which laboratory  data on solar fusion
reactions are usually fitted.
Because the WKB approximation becomes increasingly accurate
at lower energies, the standard extrapolation to solar-fusion
energies is valid.

The most compelling evidence for the
validity of the approximations of 
Eqs. (\ref{eqsfactor})--(\ref{quadapprox}) is empirical:  
they successfully fit 
low-energy laboratory data.  For example, for the 
$^3$He($^3$He,2p)$^4$He reaction, a quadratic polynomial fit
(with only a small linear and even smaller quadratic term) for
$S(E)$ provides an excellent fit to the measured cross section
over two  decades in energy in which the measured cross
section varies by over ten orders of magnitude (see discussion in
Sec.~\ref{he3he3}).

The approximation of $S(E)$ by the lowest terms in a Taylor
expansion is supported theoretically by explicit calculations
for a wide variety of reasonable nuclear potentials, for which 
$S(E)$ is found to be well approximated by a quadratic energy
dependence.
The specific form of Eq.~(\ref{eqsfactor})
describes $s$-wave tunneling through the Coulomb barrier of two
point-like nuclei.
 Several well-known and thoroughly investigated effects  
introduce slowly-varying energy dependences that are not
included explicitly in the standard definition of the low-energy 
$S$-factor. These effects include
(see, for example, Barnes, Koonin, and Langanke, 1993;\nocite{barnes93}
Descouvemont, 1993;\nocite{descouvemont93} Langanke and 
Barnes, 1996)\nocite{langanke96}
1) the finite size of the colliding nuclei,
2) nuclear structure and strong interaction effects,
3) antisymmetrization effects, 
4) contributions from other partial waves, 
5) screening by atomic electrons, and 
6) final-state phase space.
These effects introduce energy dependences in the $S$-factor that, in
the absence of near-threshold resonances, are much weaker than the
dominant energy dependence represented by the 
Gamow penetration factor. 
The standard picture of an $S$-factor with a weak energy dependence has been
found to be valid for the cross-section data of all nuclear reactions
important for the solar $pp$-chains. Theoretical energy 
dependences that take into account 
all the effects listed above are available (and 
have been used) for extrapolating 
 data for all the important reactions in solar hydrogen
burning.

One can reduce ({\it but not eliminate})
the energy dependence of the extrapolated quantity by
 removing nuclear
finite-size effects (item 1) from the data. 
The resulting modified
$\tilde S (E)$ factor is still energy dependent (because of items 2--6) 
and cannot be treated as a constant [as assumed 
by Dar and Shaviv (1996)].\nocite{Dar96}

\subsection{Laboratory Screening}
\label{labscreening}

It has generally been believed that the uncertainty in the extrapolated
 nuclear cross sections 
 is reduced by steadily lowering
the energies at which data can be taken in the laboratory. However, this
strategy has some complications (Assenbaum, Langanke, and Rolfs, 1987)
\nocite{Assenbaum} since at very low energies the
experimentally measured cross section does not represent the bare nucleus
cross section: the laboratory cross section is increased by the screening effects
arising from the electrons present in the target (and in the
projectile). The resulting enhancement of the measured cross section, 
$\sigma_{\rm exp} (E)$, relative to the cross section for bare nuclei,
$\sigma(E)$, can be written as
\begin{equation}
f(E) = \frac{\sigma_{\rm exp} (E)}{\sigma(E)}\ . 
\protect\label{eqcrosssec}
\end{equation}
Since the electron screening energy, $U_e$, is much smaller than the
scattering energies, $E$, currently accessible in experiments, one finds
(Assenbaum, Langanke, and Rolfs, 1987)\nocite{Assenbaum}
\begin{equation}
f(E) \approx {\rm exp} \left\{ \pi \eta (E) \frac{U_e}{E} \right\}\ .
\protect\label{eqfeapprox}
\end{equation}

In nuclear astrophysics, one starts with the bare
nuclei cross sections and corrects them for the screening appropriate
for the astrophysical scenario (plasma screening, 
see Sec.~\ref{stellarscreening}).  In the laboratory experiments, the 
electrons are bound to the nucleus, while in the stellar 
plasma they occupy (mainly) continuum states.
Therefore, the physical processes underlying screening effects 
 are different in the laboratory and in the plasma.

The enhancement of laboratory cross sections due to electron screening
is well established, with the ${\rm ^3He}(d,p){\rm ^4He}$ reaction being the best
studied and most convincing example (Engstler {\it et al.}, 1988; Prati
{\it et al.}, 1994)\nocite{Engstler,Prati}. However, it 
appeared for some time
that the observed enhancement was larger than the one predicted 
by theory. This discrepancy has recently been removed after improved
energy loss data became available for low-energy deuteron projectiles in
 helium gas. 
To a good approximation, atomic-target data can be corrected for
electron screening effects within the adiabatic limit (Shoppa {\it et
al.}, 1993)\nocite{Shoppa93}
in which the screening energy, $U_e$, is simply given by the difference in
electronic binding energy of the united atom and the sum of the
projectile and target atoms.  
It appears now as if the electron screening effects for
{\it atomic targets} can be  modeled reasonably well (Langanke {\it et
al.}, 1996; Bang {\it et al.}, 1996; but see also
Junker {\it et al.} 1997
)\nocite{Shoppa,Bang,junker97}. 
This conclusion must be
demonstrated for {\it molecular} and {\it solid} targets. 
Experimental work on electron screening with molecular and solid targets
is discussed in Engstler {\it et al.} (1992)\nocite{Engstler2}, 
while the first theoretical approaches are
presented in Shoppa {\it et al.} (1996)\nocite{Shoppa2} (molecular) 
and in Boudouma, Chami, and Beaumevieille (1997)\nocite{Boudouma} (solid
targets).

Electron screening effects, estimated in the adiabatic
limit, 
 are relatively small in the measured
cross sections for most solar reactions, including the important
$^3$He($\alpha,\gamma$)$^7$Be and ${\rm ^7Be}(p,\gamma){\rm ^8B}$ reactions
(Langanke, 1995)\nocite{Langanke95}. However, both 
the ${\rm ^3He(^3He},2p){\rm ^4He}$ and the 
${\rm ^{14}N}(p,\gamma){\rm ^{15}O}$ data, which extend to very  low
energies,  are enhanced due to
electron screening  and have been corrected for these effects (see
 Sec.~\ref{he3he3} and \ref{cno}).

\subsection{Stellar Screening}
\label{stellarscreening}

As shown by Salpeter (1954)\nocite{salpeter54}, the decreased 
electrostatic repulsion between 
reacting ions caused by the Debye-H\"uckel screening leads to an  
increase in reaction rates. The reaction rate enhancement factor for  
solar fusion reactions is, to an excellent approximation (Gruzinov and
Bahcall, 1998),\nocite{gruzinov98}
\begin{equation}
f = \exp \left({Z_1Z_2e^2\over kTR_D}\right),
\protect\label{eqenhancement}
\end{equation}
where $R_D$ is the Debye radius and $T$ is the temperature. 
The Debye radius is defined by the equation 
$R_D=(4\pi ne^2\zeta 
^2/kT)^{-1/2}$, where $n$ is the baryon number 
density $(\rho/m_{\rm amu})$, $\zeta = \left\{\Sigma_i X_i 
{Z^2_i\over A_i} + \left({f^\prime\over f}\right) \Sigma_i X_i {Z_i\over 
A_i}\right\}^{1/2}$, $X_i, Z_i$, and $A_i$ are, respectively, the mass fraction, 
the nuclear charge, and the atomic weight of ions of type $i$. The quantity 
$f^\prime/f \simeq 0.92$ accounts for electron degeneracy 
(Salpeter 1954)\nocite{salpeter54}. 
 Equation (\ref{eqenhancement}) is valid in the 
weak-screening limit which is defined by $kTR_D\gg Z_1Z_2e^2$. 
In the solar case, 
 screening is weak for $Z_1Z_2$ of the order 10 or less (Gruzinov and
Bahcall, 1998).\nocite{gruzinov98} Thus, plasma 
screening corrections to all important thermonuclear reaction rates are known 
with uncertainties of the order of a few percent. Although originally derived 
for  thermonuclear reactions, the Salpeter formula also describes screening 
effects on the ${\rm ^7Be}$ electron capture rate with an 
accuracy better than 1\% 
 (Gruzinov and Bahcall, 1997)\nocite{gruzinov97} (for ${\rm
^7Be}(e,\nu){\rm ^7Li}$, we 
have $Z_1=-1$, and $Z_2=4$).

Two papers questioning the validity of the Salpeter formula in 
the weak-screening 
limit appeared during the last decade, but subsequent work
demonstrated that  
the Salpeter formula was correct. The ``3/2'' controversy  
introduced by Shaviv and Shaviv (1996)\nocite{Sha} was resolved 
by Br\"uggen and Gough (1997)\nocite{Bru}; a ``dynamic screening'' 
effect discussed by Carraro, Sch\"afer, and Koonin (1988)\nocite{Car} 
 was shown to be not present by Brown and Sawyer (1997a)\nocite{brown97a} and 
Gruzinov (1997)\nocite{Gru}. 
 
Corrections of the order of a few percent to the Salpeter formula 
come from the nonlinearity of the Debye screening and from the electron 
degeneracy. There are two ways to treat these effects - numerical simulations 
 (Johnson {\it et al.}, 1992)\nocite{johnson92} and illustrative 
approximations (Dzitko {\it et al.}, 1995; Turck-Chi\`{e}ze and 
Lopes, 1993)\nocite{Dzitko}\nocite{turck93}. Fortunately, the 
asymmetry of fluctuations is not important (Gruzinov and 
Bahcall, 1997)\nocite{gruzinov97}, and numerical 
simulations of a spherically symmetrical approximation are possible even with 
nonlinear and degeneracy effects included 
(Johnson {\it et al.}, 1992)\nocite{johnson92}. The discussion of
intermediate screening by Graboske {\it et al.}
(1973)\nocite{graboske73} is not applicable to solar fusion reactions
because Graboske {\it et al.} assume complete electron degeneracy
(cf. Dzitko {\it et al.}, 1995).\nocite{Dzitko}

A fully analytical treatment of nonlinear and degeneracy effects is not 
available, but Brown and Sawyer (1997a)\nocite{brown97a} have recently 
reproduced the Salpeter 
formula by diagram summations. It would be interesting to evaluate higher order 
terms (describing deviations from the Salpeter formula) using these or similar 
methods.

\section{The \boldmath$\lowercase{pp}$ and \boldmath$\lowercase{pep}$ Reactions}
\label{pp-pep}

The rates for most  stellar nuclear reactions  are
inferred by extrapolating measurements at higher energies to stellar
reaction energies. However, the rate for the fundamental $p + p\rightarrow
 {\rm ^2D} + e^+ + \nu_e$ reaction is too small to be measured in the
laboratory.  Instead, the cross section for the $p$-$p$ reaction
 must be calculated from
standard weak interaction theory.

  The most recent calculation was performed by
Kamionkowski and Bahcall (1994),\nocite{kamionkowski} who used
improved data
on proton-proton scattering and included the
effects of vacuum polarization in a self-consistent fashion.  They
also isolated and evaluated the uncertainties
due to experimental errors and theoretical evaluations.

The calculation of the $p$-$p$ rate requires the evaluation of three
main quantities: (i) the
weak-interaction matrix element, (ii) the overlap of the $pp$
and deuteron wave functions, and (iii) mesonic exchange-current
corrections to the lowest-order axial-vector matrix element.

The best estimate for the logarithmic derivative,
\begin{equation}
     S^\prime(0) = (11.2 \pm
     0.1) \, {\rm MeV}^{-1},
\label{logderivative}
\end{equation}
is still that of Bahcall and May (1968)\nocite{bahcallmayshort}.
At the Gamow peak for the $pp$ reaction in the Sun,
this linear term provides only an ${\cal O}(1\%)$ correction to
the $E=0$ value.  The quadratic 
correction is several orders
of magnitude smaller, and therefore negligible.
Furthermore, the 1\% uncertainty in
Eq. (\ref{logderivative}) gives rise to an ${\cal O}(0.01\%)$
uncertainty in the total reaction rate.  This is negligible
compared with the uncertainties described below.  Therefore, in
the following, we focus on the $E=0$ cross-section factor.

At zero relative energy, the $S$-factor for the $pp$ reaction rate can be written
(Bahcall and May, 1968, 1969)
\nocite{bahcallmayshort,bahcallmaylong},
\begin{equation}
     S(0)=6\pi^2 m_p c \alpha \ln2\, {\Lambda^2 \over \gamma^3} \left({G_A
     \over G_V} \right)^2 {f_{pp}^R \over (ft)_{0^+ \rightarrow0^+}}
     (1+\delta)^2,
\protect\label{firstS}
\end{equation}
where $\alpha$ is the fine-structure constant; $m_p$ is the proton mass;
$G_V$ and $G_A$ are the usual Fermi and axial-vector weak coupling constants;
$\gamma=(2\mu E_d)^{1/2}=0.23161$~fm$^{-1}$ is the deuteron
binding wave number ($\mu$ is the proton-neutron
reduced mass and $E_d$ is the deuteron binding energy);
$f_{pp}^R$ is the phase-space factor for the $pp$ reaction
(Bahcall, 1966)\nocite{bahcallftau} with radiative corrections; $(ft)_{0^+
\rightarrow 0^+}$ is the $ft$
value for superallowed $0^+\rightarrow 0^+$ transitions (Savard {\it et
al.}, 1995)\nocite{savard};
$\Lambda$ is proportional to 
the overlap of the $pp$ and deuteron wave functions
in the impulse approximation (to be discussed below); and
$\delta$ takes into account mesonic corrections.

   Inserting the
current best values, we find
\begin{equation}
     S(0)=4.00\times10^{-25}\,{\rm MeV~b}\, \left({ (ft)_{0^+
     \rightarrow 0^+} \over 3073\, {\rm sec}} \right)^{-1}
     \left({\Lambda^2 \over
     6.92}\right) \left({G_A/G_V \over 1.2654} \right)^2 \left( {f_{pp}^R
     \over 0.144} \right) \left( {1+\delta \over 1.01} \right)^2.
\protect\label{firstestimate}
\end{equation}
We now discuss the best estimates and the uncertainties for each
of the factors which appear in Eq. (\ref{firstestimate}).

The quantity $\Lambda^2$ is proportional to 
the overlap of the initial-state $pp$
wave function and the final-state deuteron wave function. 
 The wave functions are determined by integrating the
Schr\"odinger equations for the two-nucleon systems with an assumed nuclear
potential.  The two-nucleon potentials cannot be determined from first
principles, but the parameters in any given functional form for the
potentials must fit the experimental data on the two-nucleon system.  By
trying a variety of dramatically different functional forms, we can
evaluate the theoretical uncertainty in the final result due to ignorance
of the form of the two-nucleon interaction.

The proton-proton wave function is obtained by solving the
Schr\"odinger equation for two protons that interact via a Coulomb plus
nuclear potential.  The potential must fit the
$pp$ scattering length and effective range determined from low-energy
$pp$ scattering.  In Kamionkowski and Bahcall (1994)\nocite{kamionkowski}, 
five forms for the
nuclear potential were considered: a square well, Gaussian,
exponential, Yukawa, and a
repulsive-core potential.  The uncertainty in $\Lambda^2$ from the
$pp$ wave function is small because there is
only a small contribution to the overlap integral from radii
less than a few fm (where the shape of the nuclear potential
affects the wave function).  At larger radii, the wave function
is determined by the measured scattering length and effective
range.  The experimental errors in the $pp$ scattering length
and effective range are negligible compared with the theoretical
uncertainties.

Similarly, the deuteron wave function must yield calculated
quantities consistent with measurements of the
static deuteron parameters, especially the binding energy, effective
range, and the asymptotic ratio of $D$- to $S$-state deuteron wave 
functions.  In Kamionkowski and Bahcall (1994)\nocite{kamionkowski}, seven 
deuteron wave functions which appear in the literature were
considered. 
The spread in $\Lambda$ due to the spread in assumed neutron-proton
interactions was 0.5\%, and the uncertainty due to experimental
error in the input parameters was negligible.

Figure~\ref{ppfigure} shows why the details of the nuclear physics
are unimportant.   
The figure displays the product of the radial $pp$ and deuteron wave
functions, $u_{pp}(r)$ and  $u_d(r)$. The wavelength of the $pp$ system is
more than an order of magnitude larger than the extent of the deuteron
wave function, so the shape of the curve shown in Fig.~\ref{ppfigure}
is independent of $pp$ energy. 
Most of the contribution to the overlap integral
between the $pp$  wave function and the
deuteron wave function comes from relatively large radii
where experimental measurements constrain the wave function
most strongly. 
The assumed shape of the nuclear potential produces
visible differences in the wave function only for
$r\lesssim 5$ fm, and these differences are small.
Furthermore, only $\sim40 $\% of the integrand comes from
$r \lesssim 5$ fm and $\sim2\%$ of the integrand comes
from $r \lesssim 2$ fm.

Including the effects of vacuum polarization and the best
available experimental parameters for the deuteron and
low-energy $pp$ scattering, one finds (Kamionkowski and Bahcall, 1994)
\nocite{kamionkowski}
\begin{equation}
     \Lambda^2 = 6.92 \times (1 \pm 0.002^{+0.014}_{-0.009}),
\protect\label{Lambdaresult}
\end{equation}
where the first uncertainty is due to experimental errors, and
the second is due to theoretical uncertainties in the form of
the nuclear potential.

An anomalously high value of $\Lambda^2=7.39$ was obtained by
Gould and Guessoum (1990),\nocite{gg} who did not make
 clear what values for the $pp$ scattering
length and effective range they used.  Even by surveying a
wide variety of nuclear potentials that fit the observed
low-energy $pp$ data, Kamionkowski and Bahcall
(1994)\nocite{kamionkowski} never
found a value of $\Lambda^2$ greater than 7.00.  We therefore
conclude that the large value of $\Lambda^2$ reported by
  Gould-Guessoum is caused by either a
numerical error or by using input data that contradict the
existing $pp$ scattering data.

The calculation of $\Lambda^2$ includes the overlap only of the
$s$-wave (i.e., orbital angular momentum $l=0$) part of the $pp$
wave function and the $S$ state of the deuteron.  
Because the matrix element is evaluated in the usual 
allowed approximation,
$D-$state components in the deuteron wave function do not contribute
to the transition.

We use $(ft)_{0^+ \rightarrow 0^+}=(3073.1\pm3.1)$,
which is the $ft$ value for superallowed
$ {0^+ \rightarrow 0^+}$ transitions that is determined from
experimental rates corrected for radiative and Coulomb effects
(Savard {\it et al.},  1995)\nocite{savard}.  
This value is obtained
from a comprehensive analysis of data on numerous 
$0^+ \rightarrow 0^+$ superallowed decays.  After radiative
corrections, the $ft$ values for all such decays are found to be
consistent within the quoted error.

Barnett {\it et al.} (1996)\nocite{pdg} recommend a 
value $G_A/G_V=1.2601\pm0.0025$,
which is a weighted average over several experiments that
determine this quantity from the neutron decay asymmetry.  However,
a recent experiment (Abele {\it et al.}, 1997)\nocite{abele} 
has obtained a slightly higher 
value.  We estimate that if we add this new result to the compilation
of  Barnett {\it et al.} (1996)\nocite{pdg}, the 
weighted average will be $G_A/G_V=1.2626 \pm
0.0033$.
Alternatively, $G_A/G_V$ may be obtained from $(ft)_{0^+
\rightarrow 0^+}$ and the neutron $ft$-value from
\begin{equation}
\left({G_A \over G_V} \right)\hskip-5pt\raise9pt\hbox{$^2$} = {1\over 3} \left[ { 2 (ft)_{0^+
     \rightarrow 0^+} \over (ft)_n } -1  \right]\ .
\protect\label{GAGV}
\end{equation}
For the neutron lifetime, we use $t_n= (888 \pm 3)$
sec.  The range spanned by this central value and the $1\sigma$
uncertainty covers the ranges given by the recommended value and
uncertainty ($887\pm2.0$) of Barnett {\it et al.} (1996)\nocite{pdg} 
and the value and
uncertainty ($889.2\pm2.2$), obtained if the results of
 Mampe (1993)\nocite{mampe}---which have been called into question by 
 Ignatovich (1995)\nocite{ignatovich}---are left out 
of the compilation.  We use the
neutron phase-space factor, $f_n=1.71465$ (including radiative
corrections), obtained in Wilkinson (1982)\nocite{wilkinson}.  Inserting the $ft$
values into Eq. (\ref{GAGV}), we find $G_A/G_V=1.2681\pm0.0033$, which
is slightly larger (by 0.0055 or 0.4\%) than the value obtained from
neutron decay distributions.  To be conservative, we take $G_A/G_V =
1.2654 \pm 0.0042$.

Considerable work has been done on corrections to the nuclear matrix
element for the exchange of $\pi$ and $\rho$ mesons
 (Gari and Huffman, 1972; Dautry, Rho, and Riska, 1976),\nocite{mesonic,dautry} which arise from nonconservation of
the axial-vector current.  By fitting an effective interaction
Lagrangian to data from tritium decay, one can show
phenomenologically that the mesonic corrections to the $pp$
reaction rate should be small (of order a few percent) (Blin-Stoyle
and Papageorgiou, 1965)\nocite{blinstoyle}.
Heuristically, this is because most of the overlap integral
comes from proton-proton separations that are large compared with the
typical ($\sim1$ fm) range of the strong interactions. 
 In tritium decay, most of the overlap of the initial and final
wave functions comes from a much smaller radius.
If mesonic effects are to be taken into account properly, they must
be included self-consistently in the nuclear potentials
inferred from data and in the calculation of the overlap
integral described above.  Here, we advocate following the
conservative recommendation of 
Bahcall and Pinsonneault (1992)\nocite{pinson} in 
adopting $\delta=0.01^{+0.02}_{-0.01}$.
The central value is consistent with the best estimates from two
recent calculations which take into account $\rho$ as well as
$\pi$ exchange (Bargholtz, 1979; Carlson {\it et al.}, 1991)
\nocite{bargholtz,carlson}.  

The quoted error range for $\delta$ could probably 
 be reduced by further investigations.  The primary 
uncertainty is not in the evaluation of exchange current matrix
elements, since the deuteron wave function is  well determined
from microscopic calculations, but in the meson-nucleon-delta
couplings that govern the strongest exchange currents.
The coupling constant combinations appearing 
in the present case are  similar to those contributing 
to tritium beta-decay, another system for which accurate microscopic
calculations can be made.  Thus the measured $^3$H lifetime 
places an important constraint on the exchange current 
contribution to the $pp$ reaction.  In the absence of a 
detailed analysis of this point, the error adopted above,
which spans the range of recently published calculations, 
remains appropriate.  But we point out that the $^3$H lifetime 
should be exploited to reduce this uncertainty.

For the phase-space factor $f_{pp}^R$, we have taken the value without
radiative corrections, $f_{pp}=0.142$ (Bahcall and May, 1969)
\nocite{bahcallmaylong} and
increased it by 1.4\% to take into account radiative corrections to
the cross section.  Although first-principle radiative corrections for this
reaction have not been performed, our best {\it ansatz}
 (Bahcall and May, 1968)\nocite{bahcallmayshort} is that they should be comparable in magnitude
to those for neutron decay (Wilkinson, 1982)\nocite{wilkinson}.  
To obtain the magnitude
of the correction for neutron decay, we simply compare the result
$f^R=1.71465$ with radiative corrections obtained in
 Wilkinson (1982)\nocite{wilkinson} to that obtained 
without radiative corrections
in Bahcall (1966)\nocite{bahcallftau}.  We estimate that the total theoretical
uncertainty in this approximation for the $pp$ phase-space factor is
0.5\%.  Therefore, we adopt $f_{pp}^R=0.144 \times (1 \pm 0.005)$,
where the error is a total theoretical uncertainty 
(see Bahcall, 1989)\nocite{bahcall89}.
It would be useful to have a first-principles calculation of
the radiative corrections for the $pp$ interaction.

Amalgamating all these results, we find that the current best estimate
for the $pp$ cross-section factor, taking account of the most recent
experimental and theoretical data, is 
\begin{equation}
     S(0)=4.00\times10^{-25}\,(1 \pm 0.007^{+0.020}_{-0.011})\,{\rm MeV~b},
\protect\label{secondbestestimate}
\end{equation}
where the first uncertainty is a $1\sigma$ experimental error, and the
second uncertainty is one-third the estimated {\it total} theoretical
uncertainty. 

Ivanov {\it et al.} (1997)\nocite{ivanov97} have recently calculated 
the $pp$ reaction rate 
 using a relativistic field theoretic model for
the deuteron.  Their calculation is invalidated by, among other
things, the fact that they used a zero-range effective interaction for the
protons, in conflict with low-energy $pp$ scattering experiments
(see Bahcall and Kamionkowski, 1997)\nocite{bk97}.

The rate for the $p+e^-+p \rightarrow {\rm ^2H} + \nu_e$
reaction is proportional to that for the $pp$ reaction.  Bahcall
and May (1969)\nocite{bahcallmaylong} found that the $pep$ 
rate could be written,
\begin{equation}
     R_{pep} \simeq 5.51 \times 10^{-5} \rho (1+X) T_6^{-1/2}
     (1+0.02 T_6) R_{pp},
\protect\label{peprate}
\end{equation}
where $\rho$ is the density in ${\rm g~cm}^{-3}$, $X$ is the mass
fraction of hydrogen, $T_6$ is the temperature in units of
$10^6$ K, and $R_{pp}$ is the $pp$ reaction rate.  This
approximation is accurate to approximately 1\% for the
temperature range, $10 < T_6 < 16$, relevant for
solar-neutrino production.  Therefore, the largest uncertainty
in the $pep$ rate comes from the uncertainty in the $pp$ rate.

\section{The $\mathbf{^3H\lowercase{e}}$($\mathbf{^3H\lowercase{e}}$, 
\boldmath$2\lowercase{p}$)$\mathbf{^4H\lowercase{e}}$ reaction }
\label{he3he3}

The solar Gamow energy of the 
${\rm^3He(^3He}, 2p){\rm^4He}$
reaction is at $E_0 = 21.4$ keV (see Eq.~\ref{defnezero}).
As early as 1972, there were desperate proposals (Fetisov and Kopysov,
1972; Fowler, 1972)\nocite{Fetisov,Fowler72} to solve the solar neutrino
problem\footnote{In 1972, the ``solar neutrino problem'' consisted
entirely of the discrepancy between the predicted and measured rates
in the Homestake experiment (see Bahcall and Davis,
1976)\nocite{bahcalldavis76}.} 
that suggested a narrow resonance 
may exist in this reaction at low energies.
Such a resonance would enhance the ${}^3{\rm He} + {}^3{\rm He}$ rate
at the expense of the ${}^3{\rm He} + {}^4{\rm He}$ chain, with important
repercussions for production of ${\rm ^7Be}$ and ${\rm ^8B}$
neutrinos.  
Many experimental investigations 
 [see Rolfs and
 Rodney (1988)\nocite{Rolfs}
for a list of references]
have searched for, but not found, an excited state 
in ${}^6{\rm Be}$ at $E_x \approx
11.6$ MeV that would correspond to a low-energy resonance in 
${}^3{\rm He} + {}^3{\rm He}$. Microscopic theoretical models 
(Descouvemont, 1994; Cs\'ot\'o, 1994)\nocite{Descouvemont,Csoto94} 
have also shown no sign of such a resonance.

Microscopic calculations of the $^3$He($^3$He,2p)$^4$He reaction 
 (Vasilevskii and Rybkin, 1989; 
Typel {\it et al.}, 1991)\nocite{Vasilevskii,Typel} view
this reaction as a two-step process: After formation of the compound
nucleus, the system decays into an $\alpha$-particle and a 2-proton
cluster. The latter, being energetically unbound, finally decays into
two protons. This, however, is expected to occur outside the range of
the nuclear forces. In Typel {\it et al.} (1991),\nocite{Typel} the 
model space was spanned by
$^4$He+$2p$ and $^3$He+$^3$He cluster functions as well as
configurations involving $^3$He pseudostates. The calculation  reproduces
the measured $S$-factors for $E \leq 300$ keV reasonably 
well and predicts $S(0) \approx 5.3$ MeV b, in agreement with the
measurements discussed later in this section.  Further confidence in the
calculated energy dependence of the low-energy $^3$He($^3$He,2p)$^4$He
cross sections is gained from a simultaneous microscopic calculation of
the analog ${\rm ^3H(^3H},2n)^4{\rm He}$ reaction, which again
 reproduces well the
measured energy dependence of the $^3$H+$^3$H fusion cross sections
 (Typel {\it et al.}, 1991)\nocite{Typel}. Recently,  
Descouvemont (1994)\nocite{Descouvemont} 
and Cs\'ot\'o (1997b, 1998)\nocite{Csoto97b,Csoto98} 
have extended the microscopic
calculations to include ${\rm^5Li}+p$ configurations. 
Their calculated energy dependences, however, are in slight disagreement
 with the data, possibly indicating the need for a genuine 3-body
 treatment of the final continuum states.

The relevant cross sections 
for the ${\rm ^3He(^3He},2p){\rm ^4He}$ reaction
have recently
been measured at the energies covering the Gamow peak.
The data have to be corrected for laboratory electron screening
effects.
Note that the extrapolation given 
by Krauss 
{\it et al.} (1987)\nocite{Krauss1} and used in 
Dar and Shaviv (1996)\nocite{Dar96} ($S(0)=5.6$ keV b) is too
high, because it is based
on low-energy data that were not corrected for electron screening.


The reaction data show that at energies below 1 MeV
the reaction proceeds predominately via a direct mechanism and
that the angular distributions approach isotropy with decreasing
energy.  The energy dependence of $\sigma(E)$---or equivalently of the
cross-section factor $S(E)$---observed by various groups 
(Bacher and 
Tombrello, 1965; Wang {\it et al.}, 1966; 
Dwarakanath and Winkler, 1971; Dwarakanath, 1974;
Krauss, Becker, Trautvetter, and Rolfs, 1987; Greife {\it et al.},
1994; Arpesella {\it et al.}, 1996; Junker {\it et al.}, 1997) 
\nocite{Bacher,Wang,Dwarakanath1,Dwarakanath2,Krauss2,Greife,%
Arpesella1,junker97} presents
a consistent picture. The only exception is the experiment
 of Good,  Kunz, and Moak (1951)\nocite{Good}, for which 
the discrepancy is most
likely caused by target problems (${}^3{\rm He}$ trapped in an Al foil).

The absolute $S(E)$ values of Dwarakanath and Winkler (1971), Krauss,
Becker, Trautvetter, and Rolfs (1987), Greife {\it et al.} (1994),
Arpesella {\it et al.} (1996), and Junker {\it et al.} (1997)  
\nocite{Dwarakanath1,Krauss2,Greife,%
Arpesella1,junker97} are in reasonable agreement, although they are
perhaps more consistent with a systematic uncertainty of $0.5$ MeV~b.
The data of 
 Wang {\it et al.} (1966)\nocite{Wang} and Dwarakanath (1974)
\nocite{Dwarakanath2}
are lower by about 25\%, suggesting a renormalization of their
absolute scales.  However, in view of the relatively few data points
reported in
 Wang {\it et al.} (1966)\nocite{Wang} and 
Dwarakanath (1974)\nocite{Dwarakanath2}, 
and their relatively large uncertainties---in comparison to 
other data sets---we suggest that the data of
 Wang {\it el al.} (1966)\nocite{Wang} and 
Dwarakanath (1974)\nocite{Dwarakanath2} can be omitted  
without significant loss of information.

Figure~\ref{datasets} is adapted from Fig. 9 of Junker {\it et al.} 
(1997)\nocite{junker97}.  The data shown are from 
 Dwarakanath and Winkler (1971), Krauss, Becker,
Trautvetter, and Rolfs (1987),  Arpesella {\it et al.}
(1996), and Junker {\it et al.} (1997)\nocite{Dwarakanath1,Krauss2,%
Arpesella1,junker97} . The data provide no evidence for a
hypothetical low-energy 
resonance over the entire energy range that has been investigated
experimentally.

Because of the effects of laboratory atomic electron screening 
 (Assenbaum, Langanke, and Rolfs, 1987)\nocite{Assenbaum},
the low-energy 
${}^3{\rm He}( {}^3{\rm He}, 2p) {}^4{\rm He}$
measurements must be corrected in order to determine the
``bare'' nuclear  $S$-factor.  Assume, for
 specificity,  a constant
laboratory screening energy of $U_e = 240$ eV, corresponding to
the adiabatic limit for a neutral ${}^3{\rm He}$ beam incident
on the atomic ${}^3{\rm He}$ target.  If we assume that the projectiles
are singly ionized, the adiabatic screening energy increases only slightly
to $U_e \approx 250$ eV.
TDHF calculations for atomic screening of low-Z targets\
 (Shoppa {\it et al.}, 1993; Shoppa {\it et al.},
 1996)\nocite{Shoppa93,Shoppa2} have shown that 
the adiabatic limit is expected to  hold well
at the low energies where screening is important. 
Junker {\it et al.} (1997)\nocite{junker97} have
converted published laboratory measurements $S_{\rm lab}(E)$ 
to bare nuclear $S$-factors
$S(E)$ using the relation 
$S(E) =  S_{\rm lab}(E) \exp(-\pi \eta(E) U_e/E)$,
with $U_e = 240$ eV [cf. Eq.~(\ref{eqcrosssec}) and Eq.~(\ref{eqfeapprox})].

The resulting bare  $S$-factors were fit 
 to Eq.~(\ref{quadapprox}).
Junker {\it et al.} (1997)\nocite{junker97}
 find $S(0) = 5.40 \pm 0.05$ MeV b, $S^\prime(0) = -4.1 \pm 0.5$ b,
and $S^{\prime^\prime}(0) = 4.6 \pm 1.0$ b$/$MeV,
but important systematic uncertainties must also be included as in
 Eq.~(\ref{S34Gpeak}) below.
An  effective $3\sigma$ uncertainty of 
about $\pm 0.30$ MeV b due to lack of
 understanding of electron screening in the laboratory experiments 
should  be
included in the error budget for $S(0)$ (cf. 
Junker {\it et al.}, 1997).

The cross-section factor at solar energies is 
relatively well known by direct measurements (see Fig.\ref{datasets}).
Junker {\it et al.} (1997)\nocite{junker97} give
\begin{equation}
S ({\rm E_0}) = 5.3 \pm 0.05{\rm(stat)} \pm 0.30{\rm (syst)}
\pm {\rm 0.30(U_e)}~{\rm MeV~b} ,
\label{S34Gpeak}
\end{equation}
where the first two quoted $1 \sigma$ errors are from statistical 
and systematic experimental uncertainties and the last error
represents a maximum likely error (or effective $3\sigma$ error) due
to 
the lack of complete understanding of laboratory electron screening.
The data seem to suggest that the effective 
value of $U_e$ may be larger than
the adiabatic limit.

Future experimental efforts should extend the $S(E)$ data to energies at
the low-energy tail of the solar Gamow peak, i.e. at least
as low as 15 keV.  Furthermore, improved data should be obtained 
at energies from E~=~25 keV to 60 keV to confirm or reject
the possibility of a relatively large systematic error 
in the $S(E)$ data near these energies.  On the theoretical
side, an improved microscopic treatment is highly desirable.

\section{The $\mathbf{^3H\lowercase{e}}$\boldmath$(\alpha,\gamma)$$\mathbf{^7B\lowercase{e}}$ Reaction}
\label{he3he4}

The relative rates of the ${\rm ^3He}(\alpha,\gamma){\rm ^7Be}$
 and ${\rm ^3He(^3He},2p){\rm ^4He}$ reactions
determine what fractions of $pp$-chain terminations result in 
${\rm ^7Be}$ or ${\rm ^8B}$ neutrinos.

Since the ${\rm ^3He}(\alpha,\gamma){\rm ^7Be}$ reaction at low
energies is essentially an external direct-capture process (Christy
and Duck, 1961), it is not surprising that direct-capture model
calculations of different sophistication yield nearly identical
energy dependences of the low-energy $S$-factor.  Both
the microscopic cluster model (Kajino and Arima, 1984)\nocite{Kajino} and the
microscopic potential model (Langanke, 1986)\nocite{Langanke} correctly predicted the
energy dependence of the low-energy ${\rm ^3H}(\alpha,\gamma){\rm
^7Li}$ cross section (the isospin mirror of ${\rm
^3He}(\alpha,\gamma){\rm ^7Be}$ ) before it was precisely measured by
Brune, Kavanagh, and Rolfs (1994).\nocite{Brune}
The absolute value of the cross section was also
predicted to an accuracy of better than $10$\% from potential model
calculations by Langanke (1986)\nocite{Langanke} and Mohr {\it et al.}
(1993).\nocite{mohr93}

Separate evaluations of this energy dependence based on the Resonating
Group Method (Kajino, Toki, and Austin, 1987)\nocite{Kajino87} 
and on a direct-capture
cluster model (Tombrello and Parker, l963)\nocite{Tombrello63} 
agree to within $\pm 1.25\%$ and are also 
in good agreement with the measured energy dependence (see also
Igamov, Tursunmuratov, and Yarmukhamedov, 1997).\nocite{igamov} 
This confluence of experiments and theory is illustrated in
 Fig.~\ref{threefour}.  Even more detailed calculations are now possible
(cf. Cs\'ot\'o, 1997a)\nocite{Csoto97a}.

Thus the energy dependence of the ${\rm ^3He}(\alpha,\gamma){\rm ^7Be}$
reaction  seems to be well determined.
The only free parameter in the extrapolation to
 thermal energies is the
normalization of the energy dependence of the cross sections 
to the measured data sets. While the energy
dependence predicted by the existing theoretical 
models is in good agreement with the energy
dependence of the measured cross sections, it would be useful to 
explore how robust this energy dependence is for a wider
range of models.
Extrapolations
based on physical models should be used; such extrapolations are more 
credible than those based only on the extension of
 multiparameter mathematical
fits (e.g., those of Castellani {\it et al.}, 1997).\nocite{Cast97}

There are six sets of 
measurements of the cross
section for the ${\rm ^3He}(\alpha,\gamma){\rm ^7Be}$ reaction that  
are based on detecting the
capture gamma rays (Table~\ref{tablehe3he4}). The weighted 
average of the six prompt $\gamma$-ray experiments yields
 a value of $S_{34}(0) = (0.507 \pm .016)$ keV~b, based on extrapolations
 made using the calculated energy dependence for this direct-capture reaction.
In computing this weighted average, we have used the renormalization
of the Kr\"awinkel {\it et al.} 
(1982)\nocite{Krawinkel82} result by 
Hilgemeier {\it et al.} (1988).\nocite{Hilgemeier88}

There are also three sets of cross sections for this reaction that are
based on measurements
of the activity of the synthesized ${\rm ^7Be}$
(Table~\ref{tablehe3he4}). 
These decay measurements
have the advantage of determining the {\it total} cross 
section directly, but
have the disadvantage that (since the source of the residual activity can
not be uniquely identified) there is always the possibility that some of the
${\rm ^7Be}$ may have been produced in a contaminant reaction 
 that evaded background tests. The three activity
measurements (when extrapolated in the same way as the direct-capture
gamma ray measurements) yield a value of $S_{34}(0) = (0.572 \pm 0.026)$
keV~b, which differs by about $2.5 \sigma$ with the value 
based on the direct-capture gamma rays. 

It has been suggested that the systematic discrepancy between these
two data sets might arise from a small monopole (E0) contribution to which
the prompt measurements would be much less 
sensitive and whose contribution could have been 
overlooked. However, estimates of the E0 contribution are consistently
found to be exceedingly small in realistic models of this reaction; 
 they are of order
$\alpha^2$, whereas the leading contribution is of order $\alpha$
(the fine structure constant).  
The importance of any E0 contributions would be
further suppressed by the fact that they would have to come 
from the $p$-wave incident channel, in contrast to the $s$-wave incident 
channel which is 
responsible for the dominant E1 contributions. (See Fig.~\ref{figcontribute}.)

When the nine experiments are combined, the weighted mean is
$S_{34}(0) = (0.533 \pm 0.013)~{\rm keV~b}$, with $\chi^2 = 13.4$
for 8 degrees of freedom.  The probability of such a distribution
arising by chance is 10\%, and that, together with the apparent
grouping of the results according to whether they have been obtained
from activation or prompt-gamma yields, suggests the possible presence
of a systematic error in one or both of the techniques.  An approach
that gives a somewhat more conservative evaluation of the uncertainty
is to form the weighted means within each of the two groups of data
(the data show no indication of non-statistical behavior within the
groups), and then determine the weighted mean of those two results.
In the absence of information about the source and magnitude of the
excess systematic error, if any, 
an arbitrary but standard prescription can be adopted in which
the uncertainties of the means of the two groups (and hence the
overall mean) are increased by a common factor of 3.7 (in this case)
to make $\chi^2 = 0.46$ for 1 degree of freedom, equivalent to making
the estimator of the weighted population variance equal to the
weighted sample variance.  The
uncertainty in the extrapolation is common to all the experiments,
although it is likely to be only a relatively 
minor contribution to the overall
uncertainty. The result is our recommended value for an
overall weighted mean:
\begin{equation}
S_{34}(0) = 0.53 \pm 0.05~{\rm keV~b} .
\label{wmean}
\end{equation}
The slope, $S^\prime(0)$,
is well determined within the accuracy of the theoretical calculations
(e.g., Parker and Rolfs,
1991):\nocite{Parker91} 
\begin{equation}
S^\prime (0) = -0.00030~{\rm b}. 
\label{slopethreefour}
\end{equation}
Neither the theoretical calculations nor the experimental data are
sufficiently accurate to determine a second derivative. 

Dar and Shaviv (1996)\nocite{Dar96} quote a value of $S_{34}(0) =
0.45$ keV~b, about $1.5\sigma$ lower than our best estimate.  The
difference between their value and our value can be traced
to the fact that Dar and Shaviv
fit the entire world set of data points as a single
group to obtain one $S_{34}(0)$ intercept, rather than 
fitting independently each
of the nine independent experiments and then combining 
their intercepts to determine a
weighted average for $S_{34}(0)$. The Dar and Shaviv method thereby
overweights the experiments of Kr\"awinkel {\it et al.} 
(1982)\nocite{Krawinkel82} and Parker and
Kavanagh (1963)\nocite{Parker63} because they have by far 
the largest number of data
points (39 and 40, respectively) and underweights those experiments
which have only 1 or 2 data points (e.g., the activity measurements).
Systematic uncertainties, such as normalization errors, common to all
the points in one data set make it impossible to treat the common points as
statistically independent and uncorrelated, and thus the Dar and Shaviv
method distorts the average.



\section{The $\mathbf{^3H\lowercase{e}}$(\boldmath$\lowercase{p}, 
\lowercase{e^+} ~+~\lowercase{\nu_e}$)$\mathbf{^4H\lowercase{e}}$ reaction }
\label{hep}

The $hep$ reaction, 

\begin{equation}
{\rm ^3He} ~+~ p ~\rightarrow ~{\rm ^4He} ~+~e^+ ~+~\nu_e ,
\label{hepreaction}
\end{equation}
produces neutrinos with an endpoint energy of $18.8$ MeV, the
highest energy expected for solar neutrinos. The region between $15$
MeV and $19$ MeV, above the endpoint energy for $^8$B neutrinos and
below the endpoint energy for $hep$ neutrinos, 
is potentially useful for solar neutrino
studies since the background in electronic detectors is expected to be
small in this energy range.
For a given solar model, the flux of $hep$ neutrinos can be calculated
accurately 
once the $S-$factor for reaction~(\ref{hepreaction}) is specified.
The rate of the $hep$ reaction is so slow that it does not affect the
solar structure calculations.
The calculated $hep$ 
flux is very small ($\sim 10^3~{\rm cm^{-2}s^{-1}}$, 
Bahcall and Pinsonneault, 1992),
but the interaction cross section is so large that the $hep$
neutrinos are potentially detectable in sensitive detectors like 
SNO and Superkamiokande (Bahcall, 1989)\nocite{bahcall89}.

The thermal neutron cross section on $^3$He has been measured
accurately in two separate experiments (Wolfs {\it et al.}, 1989; 
Wervelman, {\it et al.}, 1991).\nocite{wolfs89,wervelman}  The results are 
in good agreement with each other.

Unfortunately, there is a complicated 
 relation between the measured thermal-neutron cross
section  and the low-energy cross-section factor for the
production of $hep$ neutrinos.  The most recent detailed 
calculation (Schiavilla, {\it et al.}, 1992)\nocite{schiavilla}
 that includes $\Delta$-isobar degrees of freedom yields low-energy 
cross-section factors calculated, with specific assumptions, 
in the range $S(0) ~=~1.4\times10^{-20}$ keV~b
 to $S(0) ~=~3.2\times10^{-20}$ keV~b .
Less sophisticated calculations yield very different answers (see
 Wolfs {\it et al.},  1989; Wervelman {\it et al.}, 1991; see also the
 detailed calculation by Carlson {\it et al.}, 1991)\nocite{carlson}.

There are significant cancellations among the various
matrix elements of the one- and two-body parts of the axial current
operator.  The inferred $S$-factor is particularly sensitive to the
model for the axial exchange-current operator.  The uncertainties in the
 various components of the exchange-current operator and
the uncertainty in the weak coupling constant $g_{\beta N\Delta}$
introduce a substantial uncertainty in $S(0)$. 
Schiavilla {\it et al.}\nocite{schiavilla} use
different input parameters that reflect these
uncertainties,  and provide a range of calculated $S(0)$.

We adopt as a best estimate low-energy cross-section factor a value in
the middle of the range calculated by Schiavilla {\it et al.} (1992),
\begin{equation}
S(0)~=~2.3\times10^{-20} \, {\rm keV~b} .
\label{besthep}
\end{equation}
There is no satisfactory way of determining a rigorous error to be
associated with this best estimate.  However, we note that a factor of
$2.5$, up or down, spans the entire range of theoretical estimates that
are in the literature and therefore corresponds to the ``total
theoretical error'' often used in solar neutrino studies (Bahcall,
1989) as a
substitute for a rigorously determined $3\sigma$ uncertainty.  

Theoretical studies that could  predict  
the cross-section factor for reaction~(\ref{hepreaction}) with 
greater accuracy would be  important since
the $hep$ neutrino flux contains significant information about both
solar fusion and neutrino properties.

\section{$\mathbf{^7B\lowercase{e}}$ Electron Capture}
\label{beelectron}

The $^7$Be electron capture rate under solar conditions
has been calculated using an explicit picture of continuum-state 
and bound-state electrons and independently using a 
density matrix formulation that
does not make assumptions about the nature of the quantum states. The
two calculations are in excellent agreement within
a calculational accuracy of about 1\%.

The fluxes of both $^7$Be  and $^8$B solar neutrinos are
proportional to the ambient density of $^7$Be ions. 
The flux of $^7$Be neutrinos, $\phi({\rm ^7Be})$,
 depends upon the rate of electron capture,
$R(e)$, and the rate of proton capture, $R(p)$, 
only through the ratio
\begin{equation}
\phi({\rm ^7Be}) ~\propto~ {{R(e)} \over {R(e) + R(p)}}.
\protect\label{Beratio}
\end{equation}
With standard 
parameters, solar models yield $R(p) \approx 10^{-3} R(e)$.
Therefore, 
Equation~(\ref{Beratio}) shows that the flux of $^7$Be neutrinos
is actually independent of the local 
rates of both the electron capture and
the proton capture reactions to an 
accuracy of  better than 1\% . The $^7$Be flux depends most 
strongly on the branching between the $^3$He-$^3$He and the $^3$He-$^4$He
reactions.
The $^8$B neutrino flux is proportional to $R(p)/[R(e) + R(p)]$ and 
therefore the $^8$B flux is inversely proportional to the 
electron-capture rate.

The first detailed calculation of the $^7$Be electron capture rate
from continuum states 
under stellar conditions was by Bahcall (1962)\nocite{bahcall62}, 
who considered the thermal distribution of the electrons, the
electron-nucleus Coulomb effect, relativistic and nuclear size
corrections, and a numerical self-consistent Hartree wave 
function needed for evaluating
the electron density at the nucleus in laboratory decay (for comparison with
the electron density in stars). 
Iben, Kalata, and Schwartz (1967)\nocite{iben67} made the first explicit 
calculation 
of the effect of bound electron capture.  They included the effects of
the stellar plasma in the Debye-H\"uckel approximation and 
demonstrated that electron screening decreases significantly the bound
rate compared to the case where screening is neglected. 

Applying the same Debye-H\"uckel screening picture to
continuum states, 
Bahcall and Moeller (1969)\nocite{bahcall69}
showed  
that plasma effects on the continuum capture rate were small.
Bahcall and Moeller (1969) also formulated the total capture rate in a
convenient analytic form, which is in general use
today (Bahcall, 1989)\nocite{bahcall89},
and averaged
the capture rates over three different solar models.  Let 
$R \equiv R(e)$ be the
total capture rate and $C$ be the rate of capture from the continuum
only. Bahcall and Moeller (1969) found that the ratio of total rate to
continuum rate averaged over the solar models was 
$<R/C>  ~\simeq~<C/R>^{-1} ~=~ {1.205 \pm 0.005}$.

Watson and Salpeter (1973)\nocite{watson73} first drew attention to the small
number ($\sim 3$) of ions per Debye sphere in the solar interior; 
they emphasized the possible importance of thermal plasma fluctuations
on the bound-state electron-capture rate.  
Johnson {\it et al.} (1992)\nocite{johnson92} 
 performed a series
of detailed calculations, especially for the bound state capture rate,
using a form of self-consistent Hartree
theory. They derived a correction to the usual total rate
of about 1.4\%.

Using the previously-calculated electron capture rate as a function of
temperature, density, and composition, Bahcall
(1994)\nocite{bahcall94} 
calculated 
the fraction of decays from bound states and found that the ratio of
total to continuum captures was $R/C = 1.217 \pm 0.002$ for three modern
solar models, which is about 1\% larger than the  results of
Bahcall and Moeller (1969)\nocite{bahcall69} cited earlier.  
Using this slightly higher bound-state 
fraction, we 
find
\begin{equation}
R({\rm ^7Be} + e^-)~=~5.60\times10^{-9}(\rho/\mu_e)T_6^{-1/2}[1~+~0.004(T_6~-~16)]{\rm
s^{-1}}\ , 
\protect\label{be7ecrate}
\end{equation}
where $\mu_e$ is the electron mean molecular weight. 
In most recent discussions 
(Bahcall and Moeller, 1969; Bahcall, 1989)\nocite{bahcall69,bahcall89}, 
the numerical coefficient in Eq.~(\ref{be7ecrate})
was $5.54$ instead of $5.60$.  The slightly 
higher value given here reflects
the newer determination of the bound fraction (Bahcall, 1994)\nocite{bahcall94}.

Most recently, Gruzinov and Bahcall (1997)\nocite{gruzinov97} abandoned
the standard picture of bound and continuum states in the solar plasma
and have instead calculated the total electron capture rate directly
from the equation for the density matrix (Feynman, 1990)\nocite{feynman90} 
of the plasma.  Their
numerical results agree to within $1$\% with  the 
standard result obtained with an explicit picture of bound and
continuum electron states. They also show that a simple
heuristic argument, derivable from the density matrix formulation,
gives an analytic form for the effect of the solar plasma that is of
the familiar Salpeter (1954)\nocite{salpeter54}
form and agrees to within $1$\% with the
numerical calculations.\footnote{Even more recently, 
Brown and Sawyer (1997b)\nocite{brown97b} have
re-investigated the electron capture problem using multi-particle
field-theory methods. Their technique automatically gives the correct
weighting with Fermi statistics (a small correction) including an
account of bound states which obviates the need for ``Saha-like''
reasoning. They derive analytic sum rules which confirm the Gruzinov
and Bahcall (1997) result that the Salpeter (1954) correction holds to
good accuracy in the electron capture process.}
An explicit Monte Carlo 
calculation of the effects of
fluctuations, not required to be spherically symmetric, shows that the
net effect of fluctuations is less than $1$\% of the total capture
rate. This  result is surprising given the small number of ions in the
Debye sphere (Watson and Salpeter, 1973)\nocite{watson73}.  However, the fact that fluctuations are 
unimportant can be 
understood (or at least made plausible)
using second-order perturbation theory in the density-matrix
formulation.  The effect of fluctuations is  indeed shown (Gruzinov
and Bahcall, 1997)\nocite{gruzinov97}
 to depend
upon an inverse power ($-5/3$) of the number of ions in the Debye
sphere. But, the dimensionless coefficient is tiny ($2\times10^{-4}$).
The net result of the calculations performed with the density-matrix
formalism is to confirm to high accuracy the standard $^7$Be
electron-capture rate given here in Eq.~(\ref{be7ecrate}).

How accurate is the present theoretical capture rate, $R$? The excellent 
agreement between the numerical results obtained using different physical 
pictures (models for bound and 
continuum states and the density matrix 
formulation) suggests that the theoretical 
capture rate is relatively accurate. 
Moreover, a simple physical argument shows  
(Gruzinov and Bahcall, 1997)\nocite{gruzinov97}
 that the effects of 
electron screening on the 
total capture rate can be expressed by a Salpeter
factor (Salpeter, 1954)\nocite{salpeter54}
that yields the same numerical results as the detailed calculations. 
The  simplicity of this
 physical argument
provides supporting evidence that the calculated electron capture rate is 
robust.

The largest recognized uncertainty arises from  
possible inadequacies of the 
Debye screening theory. Johnson {\it et al.} (1992)\nocite{johnson92} 
have performed a careful 
self-consistent quantum mechanical calculation of 
the effects on the 
${\rm ^7Be}$ electron capture rate of departures from 
the Debye screening. They 
conclude that Debye screening describes the 
electron capture rates to within 2\%. 
Combining the results of Gruzinov and Bahcall (1997)\nocite{gruzinov97}  and of 
Johnson {\it et al.} (1992),  we conclude that the 
total fractional uncertainty,
 $\delta R/R$, is small and that (at about the $1\sigma$ level) 
\begin{equation}
 \frac{\delta R\left({\rm ^7Be} + e^-\right)}{R\left({\rm ^7Be} 
+ e^-\right)}\leq~0.02 .
\end{equation}
 
\section{The $\mathbf{^7B\lowercase{e}}$(\boldmath$\lowercase{p,\gamma})$
$\mathbf{^8B}$ Reaction}
\label{bep}

\subsection{Introduction}
\label{7Beintro}
The neutrino event rate in all existing solar neutrino detectors,
except for those based on the ${\rm ^{71}Ga}(\nu,e)$ reaction, 
is either dominated
by (in the case of the Homestake Mine $^{37}$Cl detector), or almost entirely
due to (in the cases of the Kamiokande, Super-Kamiokande, 
and 
SNO detectors), the high-energy neutrinos produced in $^8$B decay. 
It is therefore
important to assess critically the information needed to predict the solar
production of $^8$B.\footnote{The shape of the energy spectrum from
$^8$B decay is the same (Bahcall, 1991)\nocite{bahcalluncov91}, to one part in $10^5$, as the shape
determined by laboratory
experiments  
and is
relatively well known (see Bahcall {\it et al.},
1996)\nocite{bahcallb896}. } 
The most poorly known
quantity in the entire nucleosynthetic chain that leads to $^8$B is the
rate of the final step, the $^7$Be$(p,\gamma){\rm ^8B}$ reaction which has a
$Q$-value of $137.5 \pm 1.2$ keV (Audi and Wapstra, 1993)\nocite{au:93}.

The $^7$Be$(p,\gamma){\rm ^8B}$ rate is conventionally 
given in terms of the zero-energy
$S$-factor $S_{17}(0)$. This quantity is deduced by
extrapolating the measured absolute cross sections, which have been 
studied to 
energies as low as $E_p=134$ keV, to the astrophysically relevant 
regime.

Due to the small binding energy of $^8$B, the
${\rm ^7Be}(p,\gamma){\rm ^8B}$ reaction at low 
energies is an external, direct-capture process 
(Christy and Duck, 1961)\nocite{Christy}. Consequently 
the energy dependence of
the $S$-factor for $E \le 300$ keV is almost 
model-independent (Williams and
Koonin, 1981; Cs\'ot\'o, 1997a; Timofeyuk, Baye, and
Descouvemont, 1997)\nocite{Williams,Csoto97a,Timo97} 
and is given by the predicted ratio of
E1 capture from ${\rm ^7Be}+p$ $s$-waves and $d$-waves
into the $^8$B ground state 
(Robertson, 1973; Barker, 1980)\nocite{Robertson,Barker}. 
The $S$-factor is expected to exhibit a modest
rise at  solar energies due to  the  
energy dependences of
the Whittaker asymptotics of the ground state, the regular Coulomb
functions describing the ${\rm ^7Be}+p$ scattering states, and the 
$E_{\gamma}^3$ dipole phase-space factor. 
Because this  expected rise
of the $S$-factor towards solar energies cannot be 
seen at the energies at which capture data is currently available, 
extrapolations that do not
incorporate the correct physics of the low-energy
${\rm ^7Be}(p,\gamma){\rm ^8B}$ reaction, for
example,  the 
extrapolation presented by Dar and Shaviv (1996)\nocite{Dar96}, 
are not correct.

We have fitted Johnson {\it et al.'s} (1992) 
microscopic calculations of $S(E)$ to 
quadratic functions between 20~keV and 300~keV.  The overall
normalization was allowed to float and only the
energy dependence was fitted.  The results were practically the same for
the Minnesota force (Chwieroth {\it et al.}, 1973)\nocite{chwieroth73} and the
Hasegawa-Nagata force (Furutani {\it et al.},
1980).\nocite{furutani80}  
A combined fit,
weighting the results from both force laws equally, yields
$S^\prime(0)/S(0) = -0.7 \pm 0.2~{\rm MeV}^{-1}$ and
$S^{\prime\prime}(0)/2S(0) = 1.9 \pm 0.3~{\rm MeV}^{-2}$, which are our
recommended values.  The quadratic formulae given above reproduce the
detailed microscopic calculations to an accuracy of $\pm 0.3$~eV~b in
the energy range 0 to 300~keV.

At moderate energies, say $E \geq 400$ keV, the ${\rm
^7Be}(p,\gamma){\rm ^8B}$
$S$-factor becomes
model-dependent (e.g., Cs\'ot\'o, 1997a)\nocite{Csoto97a}, 
because at these energies the  
capture
process probes the internal $^8$B wave function and becomes  
sensitive to  nuclear structure. The argument of 
Nunes, Crespo, and Thompson (1997)\nocite{Nunes97} 
that the energy dependence of $S_{17}$ is sensitive to 
core polarization effects has been found to be invalid and the paper
has been withdrawn by the authors.
At the present time, statistical and systematic errors  in the
experimental data dominate the
uncertainty in  the low-energy cross-section factor (see also
Turck-Chi\`eze {\it et al.}, 1993).\nocite{turcketal93}  A measurement of
the cross section below 300~keV with an uncertainty significantly
better than  5\%
would make a major contribution to our knowledge of this reaction.
A  measurement of the $^7$Be quadrupole moment would also 
help distinguish between 
different nuclear models for the ${\rm
^7Be}(p,\gamma){\rm ^8B}$ reaction 
(see Cs\'ot\'o {\it et al.}, 1995).\nocite{Csoto95}


We begin by reviewing the history of direct measurements of the 
${\rm ^7Be}(p,\gamma){\rm ^8B}$ cross section. Then we 
discuss recent indirect attempts
to determine the cross section. Finally we make recommendations for
 $S_{17}(0)$.
\subsection{Direct $\mathbf{^7Be}$\boldmath($p,\gamma$)$\mathbf{^8B}$ 
measurements}
The first experimental study of ${\rm ^7Be}(p,\gamma){\rm ^8B}$ was made by 
Kavanagh (1960)\nocite{ka:60} who detected the $^8$B $\beta^+$ activity. This pioneering
measurement was followed by an experiment by 
Parker (1966, 1968),\nocite{pa:66,pa:68}
who
improved the signal-to-background by detecting the $\beta$-delayed $\alpha$'s, 
a strategy followed in all subsequent works. Subsequently, extensive
measurements were reported by Kavanagh {\it et al.}
(1969)\nocite{Kavanagh} 
in the energy region
$E_p=0.165$ to 10 MeV, and by
Vaughn {\it et al.} (1970)\nocite{va:70} at 20 
proton energies between 0.953 and 3.281 MeV. 
The most recent published works are a single point at $E_p=360$ keV by
Wiezorek {\it et al.} (1977)\nocite{wi:77} and a very 
comprehensive and careful experiment
by Filippone {\it et al.} (1983a,b),\nocite{fi:83a,fi:83b} who measured the cross section
at 25 points at center-of-mass energies between 0.117 and 1.23 MeV. 
The cross section displays a strong $J^{\pi}=1^+$ resonance at $E_p=0.72$ MeV,
but this has almost no effect at solar energies where the cross section is 
essentially due to direct $E1$ capture.

Direct ${\rm ^7Be}(p,\gamma){\rm ^8B}$ experiments require radioactive targets. It
has not been practical to use the conventional geometry with large-area, 
thin targets, and ``pencil''
beams; instead the experimenters were forced to use comparable beam and 
target sizes.
As a result the absolute normalization of the cross sections has posed
severe experimental problems. 

In the experiments to date, the mean areal density of $^7$Be atoms seen
by the proton beam has been determined in one of two ways:
\begin{enumerate}
\item counting the number of $^7$Be atoms by detecting the 478 keV photons
emitted in $^7$Be decay and measuring the target spot size
(Wiezorek {\it et al.}, 1977; Filippone {\it et al.}, 1983a,b).
\nocite{fi:83a,fi:83b}
\item measuring the yield of the ${\rm ^7Li}(d,p){\rm ^8Li}$ reaction on the 
daughter $^7$Li atoms 
that build up in the targets as the $^7$Be decays (Kavanagh, 1960;
Parker, 1966, 1968; Kavanagh {\it et al.}, 1969; Vaughn {\it et al.},
1970; Filippone {\it et al.}, 1982).\nocite{ka:60,pa:66,pa:68,Kavanagh,va:70,fi:82}  These measurements are
made on the peak of a broad ($\Gamma \approx 0.2$ MeV) 
resonance at $E_d=0.78$ MeV.
\end{enumerate}
The first method has the advantage of being direct. The second method
has the advantage that the $^8$B produced in the ($p,\gamma$) reaction
and the $^8$Li produced in the ($d,p$) calibration reaction can both be 
detected by 
counting the beta-delayed alphas, so that detection efficiency uncertainties
largely cancel out. However the second method requires an absolute measurement
of the total ${\rm ^7Li}(d,p){\rm ^8Li}$ cross section which has turned out to be rather
difficult to measure correctly.

The absolute ${\rm ^7Be}(p,\gamma){\rm ^8B}$ cross sections originally quoted from these 
experiments were not consistent with each other, 
although the shapes of the cross sections
as functions of bombarding energy were in agreement. Furthermore, the 
quoted
${\rm ^7Li}(d,p){\rm ^8Li}$ normalization cross sections also differed
by much more
than the quoted uncertainties (values differing by  up to a factor of two
were quoted). However, as pointed out by Barker and Spear (1986)\nocite{ba:86}, 
even after all the ${\rm ^7Be}(p,\gamma){\rm ^8B}$
cross sections are renormalized to a common value of the ${\rm
^7Li}(d,p){\rm ^8Li}$
cross section, the results are not consistent.

Because poorly understood systematic errors dominated the actual uncertainties
in the results, we adopt the following guidelines for
evaluating the existing data to arrive at a recommended value for $S_{17}(0)$.
\begin{enumerate}

\item We consider only those experiments that were described in sufficient 
detail that we can assess the reliability of the error assignments.

\item We review experiments that pass the above cuts and make our own 
assessment of the systematic errors, using information given in the
original  paper
plus more recent information (such as improved values for the ${\rm
^7Li}(d,p){\rm ^8Li}$
cross section) when available.
\end{enumerate}

The only low-energy ${\rm ^7Be}(p,\gamma){\rm ^8B}$ measurement 
that meets these 
criteria is the experiment 
of Filippone {\it et al.} (1983a,b)\nocite{fi:83a,fi:83b} at 
Argonne.
Filippone {\it et al.} (1983a,b)\nocite{fi:83a,fi:83b} obtained the areal density
of their target by counting the 478 keV radiation from $^7$Be
decay and also by detecting the ($d,p$) reaction on the $^7$Li produced in
the target by $^7$Be decay. The Argonne experimenters made two independent 
measurements of the ${\rm ^7Li}(d,p){\rm ^8Li}$ cross section [Elwyn {\it et al.} (1982) and 
Filippone {\it et al.} (1982)].\nocite{fi:82} These two determinations were consistent.
In addition, Filippone {\it et al.}'s (1982)\nocite{fi:82} 
gamma-ray counting and ($d,p$) normalization 
techniques gave results in excellent agreement.
\subsection{The \boldmath$^7$Li(\boldmath$d,p$)$\mathbf{^8L\lowercase{i}}$ cross 
section on the \boldmath$E = 0.6$ MeV 
resonance}
Strieder {\it et al.} (1996)\nocite{st:96} give a complete listing of 
existing ${\rm ^7Li}(d,p){\rm ^8Li}$ cross-section
measurements. The results scatter from a maximum value of $(211 \pm 15)$
mb (Parker, 1966)\nocite{pa:66} to a 
minimum of $(110 \pm 22)$ mb (Haight, Matthews,
and Bauer, 1985)\nocite{ha:85}. We obtain a recommended 
value for the ${\rm ^7Li}(d,p){\rm ^8Li}$ cross section
by applying the same criteria used above 
in evaluating the ${\rm ^7Be}(p,\gamma){\rm ^8B}$ data.
The experiments that pass our selection criteria are listed
in Table \ref{tab: (d,p) cross section}. 
The absolute cross sections given in the first 
three rows of Table \ref{tab: (d,p) cross section} 
are based on target areal densities determined from the 
energy loss of protons (McClenahan and Segal, 1975)\nocite{mc:75} or deuterons 
(Elwyn {\it et al.}, 1982 and Filippone {\it et al.},
1982)\nocite{el:82,fi:82}
 in the targets. These results therefore 
share a common
systematic uncertainty in the stopping powers. Filippone {\it et al.} 
(1982)\nocite{fi:82} cite evidence that the 
tabulated stopping powers were accurate to
5\%, but quote an overall target thickness uncertainty of 7\%. 
Elywn {\it et al.} (1982) quote a $\approx$7.5\% uncertainty in the stopping power.
McClenahan and Segal (1975) quote a target thickness uncertainty of 10\%.
 
The last two entries in Table \ref{tab: (d,p) cross section} 
differ from those given by the authors.
The next-to-last row was obtained by combining 
Filippone {\it et al.}'s (1982)\nocite{fi:82} 
two independent,
but concordant, normalizations of their target thickness. The normalization
based on counting the 478 keV photon activity from $^7$Be decay implies 
a corresponding
areal density of $^7$Li in the target, and hence can be used to give an
independent absolute normalization to their ${\rm ^7Li}(d,p){\rm ^8Li}$ 
cross section. We obtained the next-to-last value in 
Table \ref{tab: (d,p) cross section} 
by requiring that Filippone 
{\it et al.}'s (1982)\nocite{fi:82} measured
${\rm ^7Li} + d$ yield corresponded exactly to their measured $^7$Li 
areal density inferred by counting the 478 keV photons. Finally, the errors on 
the ${\rm ^7Li}(d,p){\rm ^8Li}$ cross section
quoted by Strieder {\it et al.} (1996)\nocite{st:96} are 
unrealistic. Strieder {\it
et al.} (1996)\nocite{st:96} 
used a $^7$Li beam on a D$_2$ gas target. They normalized their target
density and geometry factor to the ${\rm ^7Li} + d$ 
elastic scattering 
cross section, which they assumed had reached the Rutherford 
value at their lowest measured energy $E = 0.1$ MeV.
However, their data (see their Fig. 5) do not show that the ${\rm ^7Li}(d,p)$
cross section divided by the Rutherford cross section had become constant
at this energy. Therefore, in the last row in 
Table \ref{tab: (d,p) cross section}, 
we replace Strieder {\it et al.}'s (1996)\nocite{st:96} quoted 5\% error
in the elastic scattering cross section with an 11\% uncertainty which
is the quadratic sum of the 10\% uncertainty in the 
absolute ${\rm ^7Li}(d,p){\rm ^8Li}$
cross section quoted by Ford (1964)\nocite{fo:64} [Ford's absolute normalization 
agrees very well 
with that of Filippone {\it et al.} (1982)]\nocite{fi:82} and a 5\% uncertainty in relative
normalization of Strieder {\it et al.}'s (1996)\nocite{st:96} data to those of Ford.

We obtain our recommended value for the ${\rm ^7Li}(d,p){\rm ^8Li}$ 
cross section by the following
somewhat arbitrary procedure necessitated by the fact that McClenahan 
and Segal (1975) do not give enough information to do otherwise. 
We assume that each of the first three entries
in Table \ref{tab: (d,p) cross section} 
had assigned a 7\% uncertainty to the stopping power and subtract 
this error in quadrature from the quoted uncertainties. We then combine the 
resulting values as if they were completely independent and then add back 
a conservative 7\% common-mode error. This value is then combined with those
of the last two rows in Table \ref{tab: (d,p) cross section} 
which are treated as completely independent results.
\subsection{Indirect experiments}
Two indirect techniques have been proposed that may eventually provide 
useful quantitative
information on the low-energy ${\rm ^7Be}(p,\gamma){\rm ^8B}$ reaction: 
dissociation of $^8$B's in the Coulomb field of heavy nuclei
(Motobayashi {\it et al.}, 1994),\nocite{mo:94} and measurement
of the ${\rm ^8B}\rightarrow{\rm ^7Be}+p$ nuclear vertex constant using
single-nucleon transfer reactions
(Xu {\it et al.}, 1994).\nocite{xu:94} Motobayashi
 {\it et al.} (1994) quote a ``very preliminary value''
of $S_{17}(0)= (16.7 \pm 3.2)$ eV b.
Measurements at low bombarding energies may also
provide a constraint of $S_{17}$ (Schwarzenberg {\it et al.}, 
1996;\nocite{schwarzenberg} 
Shyam and Thompson, 1997).\nocite{shyam}

At this point it would be premature to use information from these techniques 
when deriving 
a recommended value of $S_{17}(0)$ because the quantitative validity of
the techniques has yet to be demonstrated. 

What would constitute a suitable demonstration?
In the case of the Coulomb dissociation studies,
we need a 
measurement of a dissociation reaction in which  radiative
capture can also be studied directly; the ideal 
 test case will have  many features in common with 
${\rm ^7Be}(p,\gamma){\rm ^8B}$, i.e., a low $Q$-value, 
a non-resonant $E1$ cross section,
and similar Coulomb acceleration of the reaction products.
However, 
the dissociation cross section has a very different dependence on the
multipolarity than does the radiative capture process. 
Although $^{16}$O(p,$\gamma$)$^{17}$F, $^3$H($\alpha,\gamma$)$^7$Li,
and $^{12}$C($p,\gamma$)$^{13}$N each has some of the 
desired properties, a
suitable test case in which the dominant capture multipolarity is E1 and
the nuclear structure is sufficiently simple has not yet been
identified.
On the other hand, a 
measurement of the $^{17}$F $\to$ $^{16}$O + $p$ vertex constant and the
prediction,  using the measured vertex constant, of 
 the ${\rm^{16}O}(p,\gamma){\rm^{17}F}$ capture reaction at
low energies will provide a good test of the vertex-constant technique.

To be useful as tests,
the indirect calibration reaction
and the comparison direct reaction must both be measured with an
accuracy of 10\% or better.  Otherwise, one cannot have confidence in
the method to the accuracy required for the cross section of the 
 $^7$Be($p,\gamma)^8$B reaction.
\subsection{Recommendations and conclusions}
We recommend the value
\begin{equation}
S_{17}(0)=19^{+4}_{-2}~\rm{eV~b}~,
\end{equation}
where the 1$\sigma$ error contains our best estimate of the
systematic as well as statistical errors. 
The recommended value  is based entirely on the
${\rm^7Be}(p,\gamma){\rm ^8B}$ data of Filippone 
{\it et al.} (1983a,b)\nocite{fi:83a,fi:83b} and is 
$15$\% smaller than the previous, widely-used value of $22.4$ eV~b
(Johnson {\it et al.}, 1992)\nocite{johnson92}
that was based upon a weighted average of all of the available 
experiments.
 The cross sections were obtained 
by combining Filippone {\it et al.}'s (1982)\nocite{fi:82} 
two independent determinations of the target 
areal density [for the $^7$Li$(d,p){\rm ^8Li}$ method we used the recommended 
cross section in Table \ref{tab: (d,p) cross section}], 
and extrapolated these to solar energies 
using the calculation of Johnson {\it et al.} (1992).
 It is important to note that
in the region around $E_p=1$ MeV where the two data sets overlap, 
Filippone {\it et al.}'s (1983a,b)\nocite{fi:83a,fi:83b} cross sections agree 
well with those of Vaughn 
{\it et al.} (1970).\nocite{va:70}
[We renormalized the 
Vaughn {\it et al.} (1970)\nocite{va:70} data to our recommended 
$^7$Li$(d,p){\rm ^8Li}$ cross section.] 
  
Because history has shown that the uncertainties in determining this
cross-section factor are dominated by
systematic effects, it is difficult to produce a 
3$\sigma$ confidence interval from a single acceptable measurement. 
Instead, we quote a ``prudent conservative 
range,'' outside of which it is unlikely that the ``true'' $S_{17}(0)$
lies
\begin{equation}
S_{17}(0)=19^{+8}_{-4}~\rm{eV~b}~.
\label{s17value}
\end{equation}

Past experience with measurements of the 
$^7$Be$(p,\gamma){\rm ^8B}$ cross 
section demonstrates the unsatisfactory nature of the existing situation
in which the recommended value for $S(0)$ 
depends on a single measurement. 
It is essential to have additional $^7$Be$(p,\gamma){\rm ^8B}$
measurements,  to establish  
 a secure basis for assessing the best estimate and the systematic
errors for $S_{17}(0)$.

Experiments with  $^7$Be ion beams would be 
valuable. 
Such experiments 
would avoid many of the systematic uncertainties
that are important in interpreting measurements of proton capture on a
$^7$Be target. 
For example, experiments performed with a radioactive beam can measure
the beam-target luminosity by observing the recoil protons and
Rutherford scattering.
But the ${\rm ^7Be}$-beam experiments will have their own set of  systematic
uncertainties that  must be understood.
Fortunately, experiments with  $^7$Be beams are being
initiated at several laboratories and results from the first of these
measurements may be available within a year or two.

Various theoretical calculations of the ratio of the $S$-value 
at $300$ keV and at  $20$ keV differ by several percent.  Since these 
differences will be difficult to measure, yet will contribute to the 
systematic uncertainty in future precise determinations of the solar 
$S$-value, a 
careful theoretical study should be made to try to understand the origins of 
the differences in the extrapolations.

\subsection{Late Breaking News}
\label{latebreaking}

In a recent experiment Hammache {\it et al.}  
(1998)\nocite{Hammache} measured the cross section at 14 energy points
between 0.35 and 1.4 MeV (in the center-of-mass system), excluding the
$1^+$ resonance energy range.  In this experiment two different targets
were used with different activities but similar results.  Hammache
{\it et al.} determined the ${\rm
^7Be}$ areal density using the two methods employed by Filippone {\it
et al.} (1983a,b)\nocite{fi:83a,fi:83b} 
and find consistent results.  The measured cross
section values are in excellent agreement with those of Filippone {\it
et al.}  over the wide energy range
where both experiments overlap.

Weissman {\it et al.} (1998)\nocite{weissman98} report a new 
measurement of the ${\rm ^7Li}(d,p){\rm ^8Li}$
cross section, $155 \pm 8$~mb . The
authors also draw attention to the importance of the possible 
loss of product nuclei
from the target in cross section measurements performed with 
high-Z backings.
The net result of including this new
measurement of the ${\rm ^7Li}(d,p){\rm ^8Li}$
cross section together with the values given in 
Table~\ref{tab: (d,p) cross section}, 
combined with estimates of the effect of loss of product nuclei on  
the previously computed values of $S_{17}$, 
is a cross-section factor for $^8$B production that is very close to the
best-estimate given in Eq.~(\ref{s17value}).

\section{Nuclear Reaction Rates in the CNO Cycle}
\label{cno}

The CNO reactions in the Sun form a polycycle of
reactions, among which the  main CNO-I
cycle accounts for 99\% of CNO energy production.
The contribution of the CNO cycles to the 
total solar energy output is believed
to be small, and, in standard solar models, 
CNO neutrinos account for about
2\% of the total neutrino flux. 
CNO reactions have been 
studied much less extensively than the $pp$ reactions and
therefore, in some important cases, we are unable to determine
reliable error limits for the low-energy cross-section factors.

Network calculations
show that three
reactions primarily determine the reaction rates of
the CNO cycles. 
The three reactions,
$^{14}$N$(p,\gamma)^{15}$O,  $^{16}$O$(p,\gamma)^{17}$F, and
$^{17}$O$(p,\alpha)^{14}$N, are considered in some detail in this
review.
With a nuclear reaction rate  almost 100 times slower than
the other CNO-I reactions, the
reaction ${\rm ^{14}N}(p,\gamma){\rm ^{15}O}$ 
determines, at solar temperatures, the rate of the main CNO cycle.
The $^{13}$N and $^{15}$O neutrinos
have energies and fluxes ($E_{\nu} \le 1.8$ MeV,
     $\phi_{\nu}({\rm CNO})/\phi_{\nu}({\rm ^{7}Be}) \approx 0.2$) comparable to the
$^{7}$Be neutrinos.
The production of
$^{17}$F neutrinos, with a flux two orders of magnitude smaller, is determined
by the reaction ${\rm ^{16}O}(p,\gamma){\rm ^{17}F}$
in the second cycle, while ${\rm ^{17}O}(p,\alpha){\rm ^{14}N}$ closes the
second branch of the CNO cycle.

 Figure
\ref{cycles} shows  the most important CNO reactions.

\subsection{${\rm ^{14}N}(p,\gamma){\rm ^{15}O}$}

\subsubsection{Current Status and Results}

A number of measurements of the 
$^{14}$N$(p,\gamma)^{15}$O cross section have been carried out over the past 45 years.
Most recently, Schr\"{o}der {\it et al.}
(1987)\protect\nocite{Schroder}, 
measured the prompt 
capture $\gamma$ radiations from this reaction at energies 
as low as $E_p=205$ keV; the 1957 measurements of the residual
$\beta^{+}$-activity of $^{15}$O carried out by Lamb and Hester
(1957)\protect\nocite{Lamb}  
between $E_p=100$ and 135 keV remain the lowest proton bombarding
energies to be reached in this reaction. The solar Gamow peak is
at $E_{0}=26$ keV. Three other experiments are available:  Hebbard and
Bailey (1963)\protect\nocite{Hebbard}, Pixley
(1957)\protect\nocite{Pixley},
 and Duncan and Perry (1951)\protect\nocite{Duncan}.

Table~\ref{n14s} summarizes the 
measurements and the $S$-values determined in previous publications,
as well as our recommendations.

As emphasized by Schr\"{o}der {\it et al.}
(1987)\protect\nocite{Schroder}, 
the relative contributions to the 
reaction mechanism are not fully  understood.
While Hebbard and Bailey (1963)\protect\nocite{Hebbard} analyze the data in terms of hard-sphere
direct-capture mechanisms to the ground, 6.16~MeV, and 6.79~MeV states of $^{15}$O,
Schr\"{o}der {\it et al.} (1987) find a significant contribution to the
ground-state capture
from the subthreshold resonance at $E_{R}=-504$ keV, which corresponds to the
6.79-MeV state. The agreement of the $S$-values recommended by Schr\"{o}der 
(1987)\protect\nocite{Schroder}
 and by Hebbard and Bailey (1963)\protect\nocite{Hebbard} seems
therefore accidental.
The unexplained 40\% correction to
the $\gamma$-ray detection efficiency of Schardt, Fowler, and
Lauritsen (1952)\protect\nocite{Schardt} [an experiment 
on $^{15}$N$(p,\alpha) ^{12}$C used as a
cross-section normalization by Hebbard and Bailey (1963)] and the anomalous 
energy
dependence of the cross sections in Hebbard and Bailey's (1963) analysis argue 
against
inclusion of their results in a modern evaluation of S(0).  The lack of a
refereed publication describing the work of 
Pixley (1957)\protect\nocite{Pixley}, and the use of
Geiger-counter technology in the pioneering experiment of Duncan and
Perry (1951)\protect\nocite{Duncan}, are responsible for our excluding
these data from the final evaluation.

\subsubsection{Stopping Power Corrections}

The ${\rm ^{14}N}(p,\gamma){\rm ^{15}O}$ cross sections of Lamb and
Hester (1957)\protect\nocite{Lamb} are important for our 
understanding of the CNO-I cycle,
since the data were obtained over an energy range significantly closer to the
 solar Gamow peak (about 30 keV) than other studies of this
reaction (see Table \ref{n14s}). Lamb and Hester concluded that the 
$S$-factor for
this reaction was essentially constant over the range of proton beam
energies from 100 to 135 keV with a value $S=(2.7 \pm 0.2)$ keV~b. Their
measurements were carried out using thick TiN targets and hence
measured yields were integrated over energy as the beam slowed down in
the  target.  They assumed a constant stopping
power of $2.35\times 10^{-20}$ MeV cm$^{2}$/atom, a good approximation at these
energies---a recent tabulation (Ziegler, Biersack, 
and Littmark, 1985)\nocite{Trim} gives 
values of $2.30\times 10^{-20}$ MeV
cm$^{2}$/atom at 100 keV and  $2.22\times 10^{-20}$ MeV cm$^{2}$/atom 
at 135 keV.  In
view of the intense proton beams used by Lamb and Hester, there may have been
significant hydrogen content in their targets, which would increase the
molecular stopping power by 10\% (for TiNH instead of TiN).

\subsubsection{Screening Corrections}

Low-energy laboratory fusion cross sections are enhanced by
electron screening [see Sec.~\ref{labscreening} and 
Assenbaum, Langanke, and Rolfs (1987)\protect\nocite{Assenbaum}]. 
 Screening is a
significant effect at the low energies at which Lamb and Hester 
(1957) explored the
$^{14}$N$(p,\gamma){\rm ^{15}O}$ reaction. Rolfs and Barnes
(1990)\protect\nocite{RolfsBarnes} show that screening effects 
become negligible for energy ratios $E/
U_{e}>1000$, where $U_{e}$ describes the
screening potential. This
condition is not satisfied for the data of Lamb and Hester (1957). Within 
the adiabatic approximation
(Shoppa {\it et al.}, 1993),\protect\nocite{Shoppa93} the 
screening enhancement can be estimated as
$f(E) \approx {\rm exp} \left\{ 59.6  E^{-3/2} \right\}$, with the 
scattering energy $E$ in keV.  
(This estimate has only been verified for atomic targets.)
  Screening, and the change in
the half-life of $^{15}$O from 120 to 122.2 seconds, are treated as
corrections, while the considerations related to stopping power
are viewed as included in the uncertainties quoted by Lamb and Hester. 
The screening and lifetime corrections reduce by 8\%  the
$S(0)$ value that otherwise would be inferred from the 
Lamb and Hester results.

\subsubsection{Width of the 6.79 MeV State}

Schr\"{o}der {\it et al.} (1987)\protect\nocite{Schroder} made 
detailed studies of the radiative 
capture to the
bound states of $^{15}$O, finding in one case, the ground-state transition,
marked evidence for the influence of a subthreshold state, the 6.79-MeV level.
They were  able to observe the capture to this state directly, and
 could thus obtain a proton reduced width.  
The gamma width, 
however, is
not known.  Schr\"{o}der {\it et al.} (1987) extracted 
the gamma width as a fit 
parameter, finding
an on-shell width of $6.3$
 eV.  Including the subthreshold state substantially
improves the fit to the data at energies as high as $E_p = 2.5$ MeV.
However, at the lowest energies for which the 
ground-state transition was measured, the
cross section (on the wings of the 278-keV resonance) is not
well described by the published fitting function.
Since gamma-width of the 6.79~MeV state is not well constrained,
the $S$-factor for the ground-state transition might in principle
increase even more rapidly at low energies than found by 
Schr\"{o}der {\it et al.} (1987), if the data at the lowest
measured
energies were more heavily weighted in the fitting.

Fortunately, however, there exists a precise measurement of the gamma width of
the 7.30-MeV analog state in $^{15}$N. 
Moreh, Sellyey, and Vodhanel (1981)\protect\nocite{Moreh} find for
that state that $\Gamma_\gamma$ = 1.08(8) eV, which would imply for the 6.79-MeV
state a width of 0.87 eV if analog symmetry were perfect.  An example is
known, however, of a case ($A= 13$) of an isovector E1 transition that shows
considerable departure (more than a factor of two) from analog symmetry, but a factor
of seven would be surprising.  It appears probable, therefore, that the width
of the 6.79-MeV state is not significantly larger than that found by
Schr\"{o}der
{\it et al.} (1987). 
A direct measurement of the gamma width of the 6.79~MeV state would be 
valuable.

\subsubsection{Conclusions and Recommended $S$-Factor for
${\rm ^{14}N}(p,\gamma){\rm ^{15}O}$}
\label{recS-factor}

The experiments of Schr\"{o}der {\it et al.} (1987) and  Lamb and 
Hester (1957)\protect\nocite{Lamb} can be used to estimate  
 $S(0)$ and its energy derivative.  Schr\"{o}der {\it et
al.} (1987) provide the only detailed data on the reaction
mechanism, finding that $S$ rises at lower energies as a result of the
subthreshold 
resonance at $E_R = -504$ keV, while Lamb and Hester (1957) constrain the total 
cross section at
the lowest energies. The extent to which the subthreshold resonances affect the
extrapolation to astrophysical energies is, however, limited by the known
width of
the analog state at 7.30 MeV in $^{15}$N, and, to a degree, by the total cross
section from Lamb and Hester (1957).  The value quoted by Schr\"{o}der 
{\it et al.} (1987) is
therefore  likely to represent the maximum contribution from a subthreshold
state, and cross sections could possibly range down to the values found in the
absence of the subthreshold resonance.  There is an uncertainty in the
normalization of the two experiments as well, and the overall normalization
uncertainty is derived as the quadrature of the individual
uncertainties.  

The recommended value,

\begin{equation}
S(0) = 3.5^{+0.4}_{-1.6}\    {\rm keV~b,}
\end{equation}
has been obtained by adopting the energy dependences given by Schr\"{o}der
{\it et
al.} (1987)\nocite{Schroder} in  the presence and the 
absence of the subthreshold resonance.  The
energy dependence is parameterized in terms of the intercept $S(0)$ and
$S^\prime (0)$
\begin{equation}
S^\prime (0) = -0.008[S(0) - 1.9]\  {\rm b} .
\end{equation}
The available data are insufficient to determine $S^{\prime \prime}$.

At the mean energy of 120 keV, the data of 
Lamb and Hester (1957),\nocite{Lamb} for
which the statistical and normalization uncertainty is 12\%, have been
corrected as described to give $S(120) = 2.48 \pm 0.31$ keV~b.  For  each
choice of energy dependence, those data have been converted to zero energy
and a weighted average formed with the data of 
Schr\"{o}der {\it et al.} (1987),\nocite{Schroder}
for which the statistical and normalization uncertainty is 17\%.  The
$n$-sigma upper limits on the average are a quadrature of $3.7 + n(0.45)$
and $3.2 + n(0.54)$ keV~b;  the lower limits are a quadrature of $2.5 -
n(0.30)$ and $1.9 - n(0.31)$ keV~b.  This prescription, while arbitrary,
reflects our view that the resonance and no-resonance extrapolations
represent a total theoretical uncertainty.  Hence the recommended
``three-sigma'' range is
\begin{equation}
S(0) = 3.5^{+1.0}_{-2.0} {\rm \ keV\  b.}
\end{equation}

Figure~\ref{14nfig}, adapted from Schr\"{o}der
{\it et al.} (1987), shows the extant data; the extrapolations 
shown represent the likely range of theoretical uncertainty. 
Additional uncertainty from normalization is not shown in the figure.

The uncertainty in the
$^{14}$N$(p,\gamma)^{15}$O reaction rate is much larger than
previously 
assumed, and produces comparable uncertainties in the 
calculated CNO neutrino fluxes.
On the other hand, the most important calculated solar neutrino fluxes
from the $p$-$p$ cycle are affected by at most $1$\% for a 50\% change
in the $^{14}$N$(p,\gamma)^{15}$O reaction rate,
as can be seen using  
the logarithmic partial derivatives given by Bahcall 
(1989)\nocite{bahcall89}. 

New experiments are necessary to improve the
understanding of the capture mechanism and the cross sections in
$^{14}$N$(p,\gamma)^{15}$O.

\subsection{$^{16}$O$(p,\gamma)^{17}$F}
\label{o16pgamma}

The rate of $^{17}$F neutrino production  in the Sun is 
determined primarily (see Bahcall and Ulrich, 1988)\nocite{Bah88} by the
rate of the 
$^{16}$O$(p,\gamma)^{17}$F reaction.  
A number of measurements of the
$^{16}$O$(p,\gamma)^{17}$F reaction were
made between 1949 and the early 70's, and the 
data are all in relatively good agreement.  Tanner's 
(1959)\protect\nocite{Tanner} work
is consistent with Hester, Pixley and Lamb's
(1958)\protect\nocite{Hester} 
lower-energy measurement.
Rolfs' (1973)\protect\nocite{Rolfs73} 
higher-precision work yields the value
\begin{equation}
S(0)=9.4 \pm 1.7 ~{\rm keV~b}~.
\end{equation} 
No resonance occurs below $E_{p} = 2.5$ 
MeV and a  direct capture model describes well the data over the
entire energy range studied. Since all of the experimental results are
consistent with each other, Rolfs' (1973)\nocite{Rolfs73} value is adopted.
For the latest work on this reaction, see 
Morlock {\it et al.} (1997)\nocite{morlock97}.

\subsection{$^{17}$O$(p,\alpha)^{14}$N}
\label{17O}

The  $^{17}$O$(p,\alpha)^{14}$N reaction  closes the CNO-II
branch of the CNO cycles.
The $S$-factor for this reaction has been particularly difficult to measure
or predict at solar energies, because of the large number of
resonances and the difficulty of detecting low-energy alphas.  
 Rolfs and Rodney 
(1975)\protect\nocite{Rolfs75} suggested that a 66~keV resonance may introduce
complications arising from the interference of the 5604 and 5668 keV energy
levels
of  $^{17}$O. In
1995, an experiment at Triangle Universities Nuclear Laboratory
(Blackmon {\it et al.}, 1995\protect\nocite{Blackmon})
disclosed a resonance located between 65 and 75 keV in a comparison of the 
alpha yields from $^{17}$O and $^{16}$O targets.  Experiments done by
the Bochum group (Berheide {\it et al.},
1992)\protect\nocite{Berheide}, 
on the other hand,
do not show evidence for the resonance, and exclude a resonance 
of the size seen
by Blackmon {\it et al.} (1995), but only on the basis of a smoothly varying 
background.
 The proton partial width of Blackmon {\it et al.} (1995)
 is $\Gamma_{p}=22^{+5}_{-4}$ neV while Berheide {\it et al.} 
(1992) find 
$\Gamma_{p}\le 3$ neV.  The Bochum group have recently reanalyzed 
their data, finding that a different energy calibration procedure and 
choice of background would change their upper limit to 75 neV 
(Trautvetter, 1997)\protect\nocite{privTraut}. They also 
have new radiative capture data that 
indicate an upper limit of 38 neV. Landre {\it et al.} 
(1989)\protect\nocite{Bougaert}
measured the proton
reduced width in ${\rm ^{17}O}({\rm ^3He},d){\rm ^{18}F}$, but, 
because the state is weak in proton
stripping,
uncertainties in the reaction mechanism (multi-step and compound nucleus
processes) are reflected in the uncertainty; $\Gamma_{p} = 71^{+40}_{-57}$
neV. 
We recommend using the proton width measured by Blackmon {\it et al.}
(1995), but caution the reader that contradictory data have not been
revised in the published literature.

Table \ref{170widths} summarizes the numerical results. The presence 
of a near-threshold resonance has a significant, but incompletely quantified, 
effect on the $^{17}$O$(p,\alpha)^{14}$N cross section at solar energies.

\subsection{Other CNO Reactions}
\label{othercno}

We have recomputed the cross-section factors for the ${\rm ^{12}C}
(p,\gamma){\rm ^{13}N}$ reaction, combining the data of Rolfs and
Azuma (1974)\nocite{RolfsAzuma} and Hebbard and Vogl
(1960)\nocite{HebbardVogl}.  We find $S(0) = (1.34 \pm 0.21)$~keV~b,
$S^\prime(0) = 2.6 \times 10^{-3}$~b, and $S^{\prime\prime}(0) =
8.3 \times 10^{-5}$~b/keV. For the reaction 
${\rm ^{13}C}(p,\gamma){\rm ^{14}N}$, 
we recommend the most recent
determination of the S-value reported in Table~\ref{svalues}, i.e., the
values given by  King {\it et al.} (1994)\nocite{King}.

For the ${\rm ^{15}N} (p, \alpha_0){\rm ^{12}C}$ reaction, we have
computed the weighted average cross-section factor using the results
of Redder {\it et al.} (1982)\nocite{Redder} and of Zyskind and 
Parker (1979)\nocite{Zyskind} [including the more accurate measurement
by Redder {\it et al.}  
of the cross section at the peak of the resonance].  We find a
weighted average of $S(0) = (67.5 \pm 4) \times 10^3$~keV~b. 
The cross-section derivatives are $S^\prime (0) = 310$~b and
$S^{\prime\prime}(0) = 12$~b/keV. 

For the reaction 
${\rm ^{18}O}(p,\alpha){\rm ^{15}N}$, only an approximate $S$-value 
is given since $S(E)$ cannot be 
described by the usual Taylor series and the original analysis 
by Lorenz-Wirzba {\it et al.} (1979)\protect\nocite{LorenzWirzba} 
determined directly the 
stellar reaction rates. 
Wiescher and Kettner (1982)\nocite{wiescher82} 
suggest a modification of the rate.
Very recently, Spyrou {\it et al.}
(1997)\nocite{spyrou97} have measured cross sections for the  ${\rm
^{19}F}(p,\alpha){\rm ^{16}O}$ reaction,
but the $S$-factor was not determined at energies
of interest in solar fusion.

\subsection{Summary of CNO Reactions}
\label{cnosummary}

Table \ref{svalues} summarizes the most recently published $S$-values and 
derivatives for reactions in the solar CNO-cycle. 
Since 
the reaction ${\rm ^{14}N}(p,\gamma){\rm ^{15}O}$ is the most
important for calculations of stellar energy generation and solar
neutrino fluxes, it is 
treated in detail in Table \ref{n14s} and the recommended values 
for the cross-section factor and its uncertainties are presented in
Sec.~\ref{recS-factor}.
Other CNO reactions are discussed in Sec.~\ref{o16pgamma}, 
 Sec.~\ref{17O}, and Sec.~\ref{othercno}.

\subsection{Recommended New Experiments and Calculations}

Further experimental and theoretical work on the 
${\rm^{14}N}(p,\gamma){\rm^{15}O}$ reaction is required
in order to reach the level of accuracy 
($\sim 10$\%) 
for the low-energy 
cross-section factor that is needed in stellar evolution calculations.

\subsubsection{Low-energy Cross Section}

The cross-section factor for capture directly to the ground state is
expected to increase steeply 
at
energies below the resonance energy of $278$ keV; 
direct experimental proof of this
increase is not yet available.
Experiments at the Gran Sasso underground laboratory  (LNGS) using a
 1 kg low-level Ge-detector have shown (Balysh {\it et al.},
1994)\protect\nocite{Heidelberg} no background events 
in the energy region
near $E_{\gamma}=7.5$ MeV over several days of running. A Ge-detector
arrangement coupled with a 200-kV high-current accelerator at LNGS [LUNA
phase II; Greife {\it et al.}, 1994\protect\nocite{Greife}; 
Fiorentini, Kavanagh, and Rolfs, 1995\protect\nocite{Fiorentini}; 
LUNA-Collaboration (Arpesella {\it et al.},
1996)]\protect\nocite{LUNA} would allow measurements 
down to proton energies of  
$82$ keV (corresponding to 1 event per day) and could thus confirm or
reject the predicted steep increase in $S(E)$ for 
direct captures to the ground state.
Still lower energies might be reached by detecting the $^{15}O$ residual nuclides
via their $\beta^{+}$-decay ($T_{\frac{1}{2}}=122$ s).

\subsubsection{$R$-matrix Fits and Estimates of the
$^{14}$N(p,$\gamma$)$^{15}$O Cross Section}

Though not fully described, the fit to the ground-state
transition in  Schr\"{o}der {\it et al.} (1987)
seems to be  based on single Breit-Wigner R-matrix resonances
and a direct-capture (DC) model added according to a simple prescription not
entirely consistent with  $R$-matrix theory. An alternative approach would be to
fit the groundstate transition  including direct-capture
and resonant amplitudes  following, for example, the description of Barker
and Kajino (1991)\protect\nocite{BarkerK}. 
Proper account should be taken of the target thickness. 
 Elastic scattering data of protons on
$^{14}$N
should be included in the analysis.

\subsubsection{Gamma Width Measurement of the 6.79 MeV State}

   Schr\"{o}der {\it et al.} (1987) suggest a large contribution of the 
sub-threshold
state at 6.79 MeV in  $^{15}$O to the ${\rm ^{14}N}(p,\gamma){\rm ^{15}O}$ capture
data, and find that the gamma width of that state is 6.3 eV. 
Other experiments yield only 
an upper limit of 28 fs ($\Gamma_\gamma \geq 0.024$ eV, 
Ajzenberg-Selove, 1991)\protect\nocite{Ajz} for the 
lifetime of the 6.79-MeV state. 
Depending upon the actual width,
the Variant Doppler Shift Attenuation Method 
(Warburton, Olness, and Lister, 1979\protect\nocite{Wa79}; 
Catford {\it et al.}, 1983)\protect\nocite{Ca83},
or Coulomb excitation of a $^{15}$O radioactive beam, might yield an
independent measurement of this width.
Data  on the Coulomb
dissociation of $^{15}$O could  also shed light on the
partial cross sections to the ground state 
(but not on the total cross section, which includes
important contributions from capture transitions into $^{15}$O excited
states).

\section{Discussion and Conclusions}
\label{discussion}

Table~\ref{summarytable} summarizes our best estimates, and the associated
uncertainties,  for the low-energy cross
sections of the most important solar fusion reactions. The
considerations that led to the tabulated values are discussed in
detail in the sections devoted to each reaction.

Our review of solar fusion reactions has raised a number of
questions, some of which we have resolved and others of which 
remain open and must be addressed by future measurements and
calculations.
The reader is referred to the specialized sections for a discussion of
the most important additional research that is required for each of
the reactions we discuss.

Our overall conclusion is that the knowledge of nuclear fusion
reactions under solar conditions is, in general,  detailed and
accurate and is sufficient for making relatively precise  predictions
of solar neutrino fluxes from solar model calculations.
However, a number of important steps still must be taken in order
that the full potential of solar neutrino experiments can be utilized
for astronomical purposes and for investigating possible physics
beyond the minimal standard electroweak model.

We highlight here four  of the most important reactions for which
further work is required.

$\bullet$ The only major reaction that has so far been studied in the
region of the Gamow energy peak is the $^3$He($^3$He, 2p)$^4$He
reaction.  A more detailed study of this reaction at low energies is 
required, with special attention to the region between $15$ keV and
$60$ keV.

$\bullet$ The six measurements of the $^3$He($\alpha$,$\gamma$)$^7$Be
reaction
made by direct capture differ by about $2.5\sigma$ from the measurements
made using activity measurements. Additional precision experiments
that could clarify the origin of this apparent difference would be
very valuable. It would also be important to make measurements 
of the cross section for the 
$^3$He($\alpha$,$\gamma$)$^7$Be reaction at energies 
 closer to the Gamow peak.

$\bullet$ The most important nuclear fusion reaction for interpreting
solar neutrino experiments is the $^7$Be(p,$\gamma$)$^8$B reaction.
Unfortunately, among all of the major solar fusion reactions, 
the $^7$Be(p,$\gamma$)$^8$B reaction is the 
least well known experimentally. Additional precise measurements,
particularly  at 
energies below $300$ keV,  are required in order to understand fully the
implications of the new set of solar neutrino experiments,
Super-Kamiokande, SNO, and ICARUS,  that will determine
the solar $^8$B neutrino flux with high statistical significance.

$\bullet$ The $^{14}$N(p,$\gamma$)$^{15}$O reaction 
plays the dominant role in
determining the rate of energy generation of the CNO cycle, but the rate of
this reaction is not well known.  
The most important uncertainties concern the size of the 
contribution to the total rate 
of a  subthreshold state and the absolute normalization
of the low-energy cross-section 
data. New measurements with modern techniques are required.

\section*{Acknowledgments}

This research was funded in part by the U.S. National Science
Foundation and Department of Energy.
\newpage



\newpage
\begin{table}
\tighten
\caption[]{Best estimate low-energy nuclear reaction cross-section
factors and their estimated $1\sigma$ errors.  
\protect\label{summarytable}}
\begin{tabular}{lcc}
\noalign{\smallskip}
\multicolumn{1}{c}{Reaction}&$S(0)$&$S^\prime(0)$\\
&(keV b)&(b)\\ 
\noalign{\smallskip}
\tableline
\noalign{\smallskip}
${\rm ^1H}\left(p, e^+ \nu_e\right){\rm ^2 H}$&$4.00 \left(1 \pm
0.007^{+0.020}_{-0.011}\right){\rm E-22}$&$4.48{\rm E-24}$\\
${\rm ^1H}\left(p e^-, \nu_e\right){\rm ^2 H}$&Eq. (\ref{peprate})\\
${\rm ^3He}\left({\rm ^3He}, 2p\right){\rm ^4 He}$&$\left(5.4 \pm
0.4\right) {\rm E+3}$\tablenote{Value at the Gamow peak, no derivative
required. See text for $S(0), S^\prime(0)$.} 
&\\
${\rm ^3He}\left(\alpha,\gamma\right){\rm ^7Be}$&$0.53 \pm 0.05$&$- 3.0
{\rm E-4}$\\
${\rm ^3He}\left(p, e^+ \nu_e\right){\rm ^4 He}$&$2.3{\rm E-20}$\\
${\rm ^7Be}\left(e^-, \nu_e\right){\rm ^7
Li}$&Eq. (\ref{be7ecrate})\\
${\rm ^7Be}\left(p,\gamma\right){\rm
^8B}$&$0.019^{+0.004}_{-0.002}$&See Sec. \ref{7Beintro}\\
${\rm ^{14}N}\left(p,\gamma\right){\rm
^{15}O}$&$3.5^{+0.4}_{-1.6}$&See Sec. \ref{recS-factor}
\end{tabular}
\end{table}

\begin{table}
\caption[]{Measured values of $S_{34}(0)$.\protect\label{tablehe3he4}}
\tighten
\begin{tabular}{ll}
\noalign{\smallskip}
\multicolumn{1}{c}{$S_{34}(0)$\ (keV~b)}&\multicolumn{1}{c}{Reference}\\
\noalign{\smallskip}
\tableline
\noalign{\smallskip}
\multicolumn{2}{l}{Measurement of capture $\gamma$-rays:}\\
\noalign{\smallskip}
$0.47 \pm 0.05$&Parker and Kavanagh (1963)\nocite{Parker63}\\
$0.58 \pm 0.07$&Nagatani, Dwarakanath, and Ashery (1969)\nocite{Nagatani69}
\tablenote{As extrapolated
using the direct-capture model of Tombrello and Parker (1963).\nocite{Tombrello63}}\\
$0.45 \pm 0.06$&Kr\"awinkel {\it et al.} (1982)\tablenote{As
renormalized by Hilgemeier, {\it et al.} (1988).\nocite{Hilgemeier88}}\\
$0.52 \pm 0.03$&Osborne {\it et al.} (1982, 1984)\nocite{Osborne82,Osborne84}\\
$0.47 \pm 0.04$&Alexander {\it et al.} (1984)\nocite{Alexander84}\\
$0.53 \pm 0.03$&Hilgemeier {\it et al.} (1988)\\
\multicolumn{2}{c}{Weighted Mean $= 0.507 \pm .016$}\\
\noalign{\medskip}        
\multicolumn{2}{l}{Measurement of ${\rm ^7Be}$ activity:}\\
\noalign{\smallskip}
$0.535 \pm 0.04$&Osborne {\it et al.} (1982, 1984)\\
$0.63 \pm 0.04$&Robertson {\it et al.} (1983)\nocite{Robertson83}\\
$0.56 \pm 0.03$&Volk {\it et al.} (1983)\nocite{Volk83}\\
\multicolumn{2}{c}{Weighted Mean $= 0.572 \pm .026$}\\
\end{tabular}
\end{table}

\begin{table}
\centering
\caption[]{$^7$Li$(d,p){\rm ^8Li}$ cross section ($\sigma$) at 
the peak of the $0.6$ MeV
resonance\tablenote{see also the 
discussion of Weissman {\it et al.} (1998) in 
 Sec.~\ref{latebreaking}}\protect\label{tab: (d,p) cross section}}
\begin{tabular}{lc}  
\multicolumn{1}{c}{Reference}&
$\sigma$ (mb) \\  \hline
McClenahan and Segal (1975)&$138 \pm 20$ \\
Elywn {\it et al.} (1982)&$146 \pm 13$ \\
Filippone {\it et al.} (1982)&$148 \pm 12$ \\
Filippone {\it et al.} (1982) (Our evaluation, see text)&$146 \pm 19$ \\
Strieder {\it et al.} (1996)\nocite{st:96} 
(Our evaluation, see text)&$144 \pm 15$ \\ 
\hline
Recommended value&$147 \pm 11$ \\   
\end{tabular}
\end{table}

\begin{table}
\caption[]{Cross-section factor, $S(0)$, 
for the reaction $^{14}$N$(p,\gamma)^{15}$O.
The proton energies, $E_p$, at which measurements were made are
indicated. 
\protect\label{n14s}}
\centering
\tighten
\begin{tabular}{lcr}
\multicolumn{1}{c}{$S(0)$}&$E_{p}$ &\multicolumn{1}{c}{Reference} \\
\multicolumn{1}{c}{keV~b}&MeV & \\
\hline
\noalign{\smallskip}
$3.20 \pm 0.54$ & 0.2-3.6 & Schr\"{o}der {\it et al.} (1987)
\protect\nocite{Schroder}\\
$3.32 \pm 0.12$ & &  Bahcall {\it et al.} (1982)\tablenote{Compilation
and evaluation: no original experimental data.}
\protect\nocite{Bahcall}\\
3.32 	&& Fowler, Caughlan, and Zimmerman (1975)$^{\rm a}$ 
\protect\nocite{Fowler75}\\
2.75 &0.2-1.1 &  Hebbard and Bailey (1963)\protect\nocite{Hebbard} \\
3.12  &&Caughlan and Fowler (1962)$^{\rm a}$\protect\nocite{Caughlan}\\
2.70  & 0.100-0.135 &  Lamb and Hester (1957)\protect\nocite{HesterLamb}\\
\noalign{\smallskip}
\hline
\noalign{\smallskip}
3.5$^{+0.4}_{-1.6}$ &&  Present recommended value  \\
\end{tabular}
\end{table}

\begin{table}
\caption[]{Near threshold resonance widths for ${\rm
^{17}O}(p,\alpha){\rm ^{14}N}$\protect\label{170widths}}
\centering
\tighten
\begin{tabular}{lcccr}
$^{18}$F levels (keV)	&5603.4  &5604.9 &5673    &\multicolumn{1}{c}{Reference}\\
\hline
\noalign{\smallskip}
$\Gamma_{\alpha}$ (eV)	&43 	&60 	&130 	&Mak {\it et al.} (1980), 
Silverstein {\it et al.} (1961)\protect\nocite{Mak,Silverstein}\\ 
\noalign{\smallskip}
\hline
\noalign{\smallskip}
$\Gamma_{\gamma}$ (eV)	 &0.5 	&0.9 	&1.4 	&Mak {\it et al.} (1980),
Silverstein {\it et al.}  (1961)\\
\noalign{\smallskip}
\hline
\noalign{\smallskip}
	&	&	&71$^{+40}_{-57}$  &Landre {\it 
at al.} (1989)\protect\nocite{Bougaert}\\ 
$\Gamma_{p}$ (neV)	&	&	&$\le$3, $\le$75 	&Berheide 
{\it et al.} (1992)\protect\nocite{Berheide}\\
	&	&	&22$^{+5}_{-4}$    &Blackmon {\it 
et al.} (1995)\protect\nocite{Blackmon}\\ 
\end{tabular}
\end{table}

\begin{table}
\tighten
\caption{Summary of published $S$-values and derivatives for  CNO 
reactions. See text for details and discussion.
When more than one $S$-value is given, the recommended value is indicated
in the table.\protect\label{svalues}} 
\medskip
\begin{tabular}{lllccr}  
\multicolumn{1}{c}{Reaction}&\multicolumn{1}{c}{Cycle}
&\multicolumn{1}{c}{$S(0)$}&$S^\prime(0)$&$S^{\prime\prime}(0)$
&\multicolumn{1}{c}{Reference}\\ 
&	 &\multicolumn{1}{c}{keV~b}	&b	&b~keV$^{-1}$	& \\
\noalign{\smallskip}
\hline
\noalign{\smallskip}
${\rm ^{12}C}(p,\gamma){\rm ^{13}N}$ &I&$1.34 \pm
0.21$&2.6E$-$3&8.3E$-$5&Recommended; this paper\\ 
&&1.43 & & & Rolfs and Azuma (1974)\protect\nocite{RolfsAzuma}\\
& &$1.24 \pm 0.15$ & & &Hebbard and Vogl (1960)\protect\nocite{HebbardVogl}\\
\noalign{\smallskip}
 \hline
\noalign{\smallskip}
${\rm ^{13}C}(p,\gamma){\rm ^{14}N}$ &I &$7.6 \pm 1$
&\llap{$-$}7.8E$-$3 &7.3E$-$4 
&Recommended; King {\it et al.} (1994)\protect\nocite{King}\\ 
& &$10.6 \pm 0.15$  & & & Hester and Lamb (1961)\protect\nocite{HesterLamb}\\
&&$5.7 \pm 0.8$&&&Hebbard and Vogl (1960)\nocite{HebbardVogl}\\
& &8.2  & & & Woodbury and Fowler (1952)\protect\nocite{Woodbury}\\
\noalign{\smallskip}
\hline
\noalign{\smallskip}
${\rm ^{14}N}(p,\gamma){\rm ^{15}O}$ &I &$3.5^{+0.4}_{-1.6}$&see text &  & see Table~\ref{n14s}     \\
\noalign{\smallskip}
\hline
\noalign{\smallskip}
${\rm ^{15}N}(p, \alpha_{0}){\rm ^{12}C}$ &I 
&($6.75 \pm 0.4$)E+4 &310 &12 
&Recommended; this paper\\
& &($6.5 \pm 0.4$)E+4 & & 
&Redder {\it et al.} (1982)\protect\nocite{Redder}\\
& &($7.5 \pm 0.7$)E+4 &351  &11 & Zyskind and 
Parker (1979)\protect\nocite{Zyskind}\\
& &5.7E+4 & & &Schardt, Fowler, and Lauritsen (1952)\protect\nocite{Schardt}\\
\noalign{\smallskip}
\hline
\noalign{\smallskip}
${\rm ^{15}N}(p, \alpha_{1}){\rm ^{12}C}$ &I &0.1 & & 
&Rolfs (1977)\protect\nocite{Rolfs77}\\
\noalign{\smallskip}
\hline
\noalign{\smallskip}
${\rm ^{15}N}(p,\gamma){\rm ^{16}O}$ &II  &$64 \pm 6$ &2.1E$-$2 &4.1E$-$3
&Rolfs and Rodney (1974)\nocite{Rolfs74}\\
\noalign{\smallskip}
\hline
\noalign{\smallskip}
${\rm ^{16}O}(p,\gamma){\rm ^{17}F}$ &II &$9.4 \pm 1.7$ &\llap{$-$}2.4E$-$2 
&5.7E$-$5&    Rolfs (1973)\\
\noalign{\smallskip}
\hline
\noalign{\smallskip}
${\rm ^{17}O}(p, \alpha){\rm ^{14}N}$ &II &
& & &Brown (1962) (see Table~\ref{170widths}) \protect\nocite{Br62}\\
&&&&&Kieser, Azuma, and Jackson 
(1979)\protect\nocite{Kieser}      \\ 
\noalign{\smallskip}
\hline
\noalign{\smallskip}
${\rm ^{17}O}(p,\gamma){\rm ^{18}F}$ &III  &$12 \pm 2$ & &   &Rolfs (1973) 
\\ 
\noalign{\smallskip}
\hline
\noalign{\smallskip}
${\rm ^{18}O}(p, \alpha){\rm ^{15}N}$ &III &$\sim 4$E+4 & & 
&Lorenz-Wirzba {\it et al.} (1979)\protect\nocite{LorenzWirzba}      \\
\noalign{\smallskip}
\hline
\noalign{\smallskip}
${\rm ^{18}O}(p,\gamma){\rm ^{19}F}$ &IV &$15.7 \pm 2.1$ &3.4E$-$4 
&$-2.4$E$-$6   
&Wiescher {\it et al.} (1980)\protect\nocite{Wiescher}\\ 
\end{tabular}
\end{table}
\newpage

\begin{figure}
\tightenlines
\caption[]{The figure shows the integrand, $u_{pp}(r) \times u_d(r)$,
of the nuclear matrix element $\Lambda$ versus radius (fm). The
ordinate is given in units of (fm)$^{-1/2}$.
Here $u_{pp}(r)$ and $u_d(r)$
are, respectively, the radial wave functions of the $p$-$p$ initial
state and the deuteron final state.  The figure (taken from
Kamionkowski and Bahcall, 1994)\nocite{kamionkowski} displays the
integrand calculated assuming five very different $p$-$p$ potentials.
In (a) we show the overlap out to a radius of $50\ $fm, while in (b)
we magnify the first $5\  $fm.  
Even drastic changes in the $p$-$p$
potential result in relatively small changes of the integrand.
\protect\label{ppfigure}}
\bigskip

\caption[]{This figure is adapted from Fig.~9 of Junker {\it et
al.} (1997), a recent paper by the LUNA Collaboration.  The measured
cross-section factor $S(E)$ for the ${\rm ^3He (^3He}, 2p){\rm ^4He}$ reaction is
shown and a fit with a screening potential $U_e$ is illustrated.  The
Gamow peak at the solar central temperature is shown in arbitrary units.
The data shown here correspond to a bare nucleus value at zero energy 
of $S(0) = 5.4$ MeV~b and a value at the 
Gamow peak of $S({\rm Gamow~Peak}) = 5.3$ MeV~b. 
\protect\label{datasets}}
\bigskip

\caption[]{Comparison of the energy dependence of the direct-capture
model calculation (Tombrello and Parker, 1963)\nocite{Tombrello63}
with the energy dependence of each of the four $S_{34}(E)$ data sets
which cover a significant energy range.  The data sets have been
shifted arbitrarily in order to show the comparison of the calculation
with {\it each} data set.\hfil\break
[Hi88]: (Hilgemeier {\it et al.}, 1988)\nocite{Hilgemeier88}\hfil\break
[Kr82]: (Kr\"awinkel {\it et al.}, 1982)\nocite{Krawinkel82}\hfil\break
[Os82]: (Osborne {\it et al.}, 1982)\nocite{Osborne82}\hfil\break
[Pa63]: (Parker and Kavanagh, 1963)\nocite{Parker63}\hfil\break
\protect\label{threefour}}
\bigskip

\caption[]{Model calculations (Tombrello and Parker, 1963) of the fractional
contributions of various partial waves and multipolarities to the total
(ground state plus first excited state) 
${\rm ^3He}(\alpha,\gamma){\rm ^7Be}$ direct-capture cross section factor.
\protect\label{figcontribute}}
\bigskip

\caption[]{CNO reactions summarized in schematic form.  The widths of the arrows
illustrate the significance of the reactions in determining the nuclear 
fusion rates in the solar CNO cycle.  Certain 
``Hot CNO'' processes are indicated by dotted
lines.\protect\label{cycles}}
\bigskip

\caption[]{Cross sections for ${\rm ^{14}N}\left(p,\gamma\right){\rm ^{15}O}$, expressed as $S(E)$, from 
extant experimental data. The
data of Lamb and Hester \nocite{Lamb} have been corrected as described in the
text.  The curves represent the low-energy extrapolations that would be obtained
under the two assumptions of no subthreshold resonance (dotted) at
$E_R = -504$ keV, and a
resonance of the strength considered by Schr\"{o}der {\it et al.} 
(dashed).\protect\label{14nfig}}
\end{figure}
\newpage
\end{document}